\documentclass[aps,prd,superscriptaddress,
showpacs,preprintnumbers]{revtex4}

\usepackage{graphicx,color,here}

\newcommand{\be}{\begin{equation}}
\newcommand{\ee}{\end{equation}}
\newcommand{\bea}{\begin{eqnarray}}
\newcommand{\eea}{\end{eqnarray}}

\newcommand{\xbj}{x_{\!\scriptscriptstyle B}}

\newcommand{\bfk}{\mbox{\boldmath $k$}}
\newcommand{\bfq}{\mbox{\boldmath $q$}}
\newcommand{\bfP}{\mbox{\boldmath $P$}}
\newcommand{\bfS}{{\mbox{\boldmath $S$}}_{_T}}
\newcommand{\bfl}{\mbox{\boldmath $\ell$}}
\newcommand{\pup}{p^\uparrow}
\newcommand{\pdown}{p^\downarrow}

\newcommand{\bfp}{\mbox{\boldmath $p$}}

\def\lsim{\mathrel{\rlap{\lower4pt\hbox{\hskip1pt$\sim$}}\raise1pt\hbox{$<$}}}
\def\gsim{\mathrel{\rlap{\lower4pt\hbox{\hskip1pt$\sim$}}\raise1pt\hbox{$>$}}}
\def\nostrocostruttino#1\over#2{\mathrel{\mathop{\kern 0pt \rlap
{\hbox{$#1$}}} \hbox{\kern-.135em $#2$}}}

\newcommand{\NP}[1]{{\it Nucl.\ Phys.}\ {\bf #1}}
\newcommand{\ZP}[1]{{\it Z.\ Phys.}\ {\bf #1}}
\newcommand{\PL}[1]{{\it Phys.\ Lett.}\ {\bf #1}}
\newcommand{\PR}[1]{{\it Phys.\ Rev.}\ {\bf #1}}
\newcommand{\PRL}[1]{{\it Phys.\ Rev.\ Lett.}\ {\bf #1}}

\def\kt{k_\perp}

\def\ptq{p_\perp}
\def\pt{P_T}
\begin{document}

\title{The role of Cahn and Sivers effects in Deep Inelastic Scattering}

\author{M.~Anselmino}
\affiliation{Dipartimento di Fisica Teorica, Universit\`a di Torino and
          INFN, Sezione di Torino, Via P. Giuria 1, I-10125 Torino, Italy}
\author{M.~Boglione}
\affiliation{Dipartimento di Fisica Teorica, Universit\`a di Torino and
          INFN, Sezione di Torino, Via P. Giuria 1, I-10125 Torino, Italy}
\author{U.~D'Alesio}
\affiliation{INFN, Sezione di Cagliari and Dipartimento di Fisica,
Universit\`a di Cagliari, C.P. 170, I-09042 Monserrato (CA), Italy }
\author{A.~Kotzinian}
\affiliation{Dipartimento di Fisica Generale, Universit\`a di Torino and
          INFN, Sezione di Torino, Via P. Giuria 1, I-10125 Torino, Italy\\
          Yerevan Physics Institute, Alikhanian Brothers St. 2; AM-375036
          Yerevan, Armenia; \\
          JINR, Dubna, 141980, Russia}

\author{F.~Murgia}
\affiliation{INFN, Sezione di Cagliari and Dipartimento di Fisica,
Universit\`a di Cagliari, C.P. 170, I-09042 Monserrato (CA), Italy }
\author{A.~Prokudin}
\affiliation{Dipartimento di Fisica Teorica, Universit\`a di Torino and
          INFN, Sezione di Torino, Via P. Giuria 1, I-10125 Torino, Italy}

\begin{abstract}
\noindent
The role of intrinsic $\bfk_\perp$ in inclusive and semi-inclusive Deep
Inelastic Scattering processes ($\ell \, p \to \ell \, h \, X$) is studied
with exact kinematics within QCD parton model at leading order; the dependence
of the unpolarized cross section on the azimuthal angle between the leptonic
and the hadron production planes (Cahn effect) is compared with data and used
to estimate the average values of $k_\perp$ both in quark distribution and
fragmentation functions. The resulting picture is applied to the description
of the weighted single spin asymmetry $A_{UT}^{\sin(\phi_\pi - \phi_S)}$
recently measured by the HERMES collaboration at DESY; this allows to extract
some simple models for the quark Sivers functions. These are compared with the
Sivers functions which succeed in describing the data on transverse single
spin asymmetries in $\pup \, p \to \pi \, X$ processes; the two sets of
functions are not inconsistent. The extracted Sivers functions give
predictions for the COMPASS measurement of $A_{UT}^{\sin(\phi_\pi - \phi_S)}$
in agreement with recent preliminary data, while their contribution to HERMES
$A_{UL}^{\sin\phi_\pi}$ is computed and found to be small. Predictions for
$A_{UT}^{\sin(\phi_K - \phi_S)}$ for kaon production at HERMES are also given.
\end{abstract}
\maketitle
\section{Introduction}

It is becoming increasingly clear that unintegrated parton distribution
and fragmentation functions play a significant role in many physical processes 
and that they are more fundamental objects than the usual integrated versions;
intrinsic $\bfk_\perp$ originates both from partonic confinement and from
basic QCD evolution \cite{ab} and often cannot be ignored in perturbative
QCD hard processes and in soft non perturbative physics. For example, this
was already pointed out in Refs. \cite{fff} concerning the computation
of the unpolarized cross section for inclusive hard scattering processes
like $p \, p \to h \, X$ at intermediate energy values. Similarly, in
semi-inclusive Deep Inelastic Scattering (SIDIS) processes,
$\ell \, p \to \ell \, h \, X$, the intrinsic partonic motion results in an
azimuthal dependence of the produced hadron $h$ \cite{cahn, kk, EMC1, EMC2}.
A full consistent treatment of several inclusive $p\,p$ processes with all
intrinsic motions, in different kinematical regions, has been recently
discussed in Ref. \cite{fu}.

While the role of intrinsic $\bfk_\perp$ can be important in
unpolarized processes, it becomes crucial for the explanation of
many single spin effects recently observed and still under active
investigation in several ongoing experiments; spin and
$\bfk_\perp$ dependences can couple in parton distributions and
fragmentations \cite{rev}, thus giving origin to unexpected
effects in polarization observables. One such example is the
azimuthal asymmetry observed by the HERMES collaboration in the
scattering of unpolarized leptons off polarized protons
\cite{hermUL, hermUT}. Another striking case is the observation of
large transverse single spin asymmetries (SSA) in $\pup \, p \to
\pi \, X$ processes \cite{e704, star}.

We consider here in a consistent approach the role of parton intrinsic 
motion in inclusive and semi-inclusive DIS processes within QCD parton 
model at leading order; the kinematics of intrinsic $\bfk_\perp$ is fully 
taken into account in quark distribution functions, in the elementary 
processes and in the quark fragmentation process. This induces several 
corrections of order $\kt/Q$ or higher, which are exactly computed: however, 
we do not consider similar corrections which might originate from higher-twist 
distribution and fragmentation functions \cite{mt}. 
The average values of $k_\perp$ for quarks inside 
protons, and for final hadrons inside the fragmenting quark jet, are fixed by 
a comparison with data on the dependence of the unpolarized cross section 
on the azimuthal angle between the leptonic and the hadronic planes. 

Such values are then used to compute the SSA for
$\ell \, \pup \to \ell \, \pi \, X$ processes, which would be zero without any
intrinsic motion. We concentrate on the Sivers mechanism \cite{siv}, that is
the spin and $\bfk_\perp$ dependence in the distribution of unpolarized quarks
inside a transversely polarized proton; it can be isolated by
studying the weighted SSA $A_{UT}^{\sin(\phi_\pi - \phi_S)}$ \cite{bm},
recently measured by HERMES \cite{hermUT} and, still preliminarily, by
COMPASS \cite{compUT} collaborations. Although the data are still scarce,
with large errors, a first qualitative estimate of the quark Sivers functions
can be obtained; the information gathered from HERMES data is in agreement
with the preliminary COMPASS data. The contribution of these functions to
the weighted longitudinal SSA $A_{UL}^{\sin\phi_\pi}$ is computed and shown to
be negligible: the measured $A_{UL}^{\sin\phi_\pi}$ \cite{hermUL} can be
equally originated not only by the Sivers mechanism, but also by the Collins
effect \cite{col}, occurring in the fragmentation of a transversely polarized
quark, and by higher-twist contributions.

A similar analysis of SSA in $\pup \, p \to \pi \, X$ processes, with a
separate study of the Sivers and the Collins contributions, has been performed
respectively in Refs. \cite{fu} and \cite{noicol}, with the conclusion
that the Sivers mechanism alone can explain the data \cite{e704}, while the
Collins mechanism is strongly suppressed. Explicit expressions of the quark
Sivers functions, as obtained from data on SSA in $\pup \, p \to \pi \, X$
processes, are given in ref. \cite{fu}. They are qualitatively similar to
the Sivers functions obtained here from SIDIS processes. Let us notice
that the universality of the Sivers function is an important open issue;
it has been proven that the quark Sivers functions in SIDIS and
in Drell-Yan processes must be opposite \cite{colsdy}, while no definite
conclusion has been theoretically reached concerning the relation between
the Sivers function in SIDIS and $pp$ processes \cite{cm}.

\section{Definitions and kinematics}

\begin{figure}[t]
\begin{center}
\scalebox{0.4}{\input{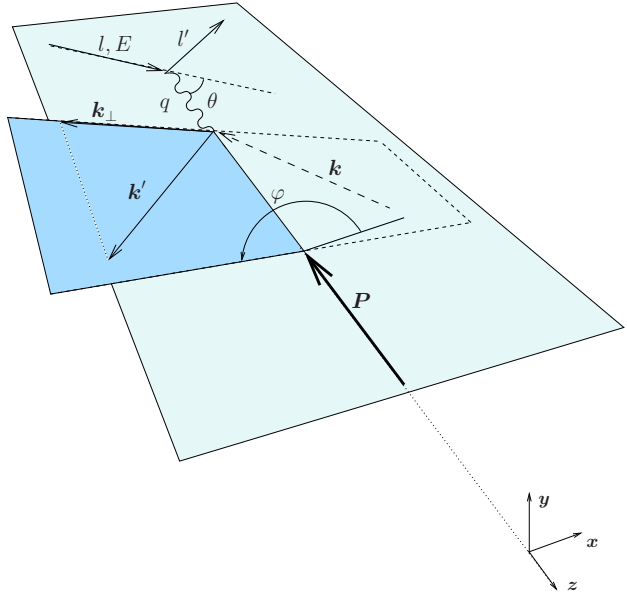}}
\caption{\small Kinematics of  DIS processes.}
\label{fig:planesdis}
\end{center}
\end{figure}

Let us start from the kinematics of Deep Inelastic Scattering processes
in the $\gamma^* p$ c.m. frame, as shown in Fig. \ref{fig:planesdis}.
We take the photon and the proton colliding along the $z$ axis with momenta
$\bfq$ and $\bfP$ respectively; the leptonic plane coincides with the
$x$-$z$ plane (following the so-called ``Trento conventions'' \cite{trento}).
We adopt the usual DIS variables (neglecting the lepton mass):
\bea
s = (P + \ell)^2 \quad\quad  Q^2 = -q^2 \quad\quad
(P+q)^2 = W^2 = \frac{1-\xbj}{\xbj}\,Q^2 + m_p^2 \nonumber \\
\xbj = \frac {Q^2}{2P \cdot q} = \frac{Q^2}{W^2 + Q^2 - m_p^2}
\quad\quad y = \frac{P \cdot q}{P \cdot \ell} = \frac{Q^2}{\xbj (s - m_p^2)}
\> \cdot \label{kin}
\eea
%
%
%

If one neglects also the proton mass $m_p$ the
four-momenta involved can be written as:
\bea
\nonumber
\ell &=& E(1,\sin\theta,0,\cos\theta) = (E, \bfl)\\
\nonumber
\ell ^\prime &=& \ell - q \\
\nonumber
q &=& \frac{1}{2}\left ( W - \frac{Q^2}{W},0,0, W + \frac{Q^2}{W}\right )
\label{eq:kinematics} \\
P &=& P_0(1,0,0,-1) \;, \nonumber
\eea
where
\bea
&& \cos \theta = \frac{Q^2 + E(W-Q^2/W)}{E(W+Q^2/W)} =
\frac{1+ (y-2)\,\xbj}{1-y\,\xbj} \nonumber \\
&& \quad\quad E = \frac{s - Q^2}{2W}
\quad\quad P_0 = \frac{1}{2}\left ( W + \frac{Q^2}{W} \right )
\> \cdot
\eea

At leading QCD order the lepton scatters off a quark and, taking intrinsic
motion into account, the initial and final quark four-momenta are given by:
\be
k = \left (xP_0+\frac{\kt^2}{4xP_0},\,\bfk _\perp,\,
-xP_0+\frac{\kt^2}{4xP_0}\right ) \quad\quad\quad
k^\prime = k+q
\ee
where $x = k^-/P^-$ is the light-cone fraction of the proton momentum carried
by the parton and $\bfk _\perp = \kt(\cos\varphi,\sin\varphi,0)$ is the
parton transverse momentum, with $\kt \equiv |\bfk _\perp|$.

The elementary Mandelstam variables $\hat s = (l+k)^2$, $\hat t = (l-l')^2$,
and $\hat u = (k-l')^2$ are then given by:
\bea
\hat s &=& xs - 2 \bfl \cdot \bfk_\perp - \kt^2 \frac{\xbj}{x}
\left( 1 - \frac{\xbj s}{Q^2} \right) \nonumber \\
\hat t &=& - Q^2 \label{stu} \\
\hat u &=& - x \left( s - \frac{Q^2}{\xbj} \right) + 2 \bfl \cdot \bfk_\perp
- \kt^2 \frac{\xbj^2 s}{x Q^2} \> \cdot \nonumber
\eea
The on-shell condition for the final quark
\be
k'^2 = 2 q \cdot k - Q^2 = \hat s + \hat t + \hat u = 0
\ee
implies
\be
x = \frac{1}{2} \, \xbj\left ( 1+ \sqrt{1+\frac{4\kt^2}{Q^2}}\right ) \cdot
\label{x}
\ee

Notice that when terms ${\cal O}(\kt^2/Q^2)$ are neglected in the above
equations one recovers the usual relations $x=\xbj$ and
$\bfk = \xbj\bfP + \bfk _\perp$; however, a significant dependence on the
azimuthal angle $\varphi$ remains, at ${\cal O}(\kt/Q)$, in the
partonic Mandelstam invariants $\hat s$ and $\hat u$, via
\be
\bfl \cdot \bfk_\perp = E \kt \, \sin\theta \, \cos\varphi \>. \label{azi}
\ee
One explicitly has
\bea
\hat s^2 &=& \frac{Q^4}{y^2} \left[ 1 - 4 \frac{\kt}{Q}\, \sqrt{1-y} \,
\cos\varphi \right] + {\cal O} \left( \frac{\kt^2}{Q^2} \right)
\label{eq:s2}\\
\hat u^2 &=& \frac{Q^4}{y^2} \, (1-y)^2 \left[ 1 - 4 \frac{\kt}{Q} \,
\frac{\cos\varphi}{\sqrt{1-y}} \right] +
{\cal O} \left( \frac{\kt^2}{Q^2} \right) \cdot
\label{eq:u2}
\eea

\subsection{DIS cross section \label{section-dis}}

Let us recall the expression of the cross section for Deep Inelastic
Scattering processes, $\ell \, p \to \ell \, X$, in the framework of the QCD
parton model with the inclusion of intrinsic $\bfk _\perp$ \cite{lp, aram}.
One starts from
\be
\frac{ d^2 \sigma ^{\ell p\to \ell X}}{d \xbj \, d Q^2} =
\frac{\pi \alpha^2}{Q^2}\frac{1}{\xbj^2 \, s^2} \,
L_{\mu\nu}W^{\mu\nu}\;,
\label{dsig}
\ee
with
\be
L_{\mu\nu} = 2 (l_\mu l'_\nu + l'_\mu l_\nu - g_{\mu\nu} \, l\cdot l')\;,
\label{eq:lmunu}
\ee
and
\be W^{\mu\nu} = \sum_q \int dx \, d^2 \bfk _\perp  \left (
\frac{1}{x}\right ) f_q(x, k_\perp) \, w^{\mu\nu} \;,
\label{eq:Wmunu1} \ee where $f_q(x, k_\perp)$ is the number
density of quarks of flavor $q$ inside the initial hadron,
carrying a transverse momentum $\bfk_\perp$ and a light-cone
fraction $x$ of the proton momentum. The elementary quark tensor
is given by
\be
w_{\mu\nu}  = 2 e_q^2 \left (k_\mu k'_\nu + k'_\mu k_\nu - g_{\mu\nu}
\frac{Q^2}{2}\right ) \delta ( 2q\cdot k -Q^2)\;,
\label{eq:wmunu}
\ee
so that
\be
L_{\mu\nu} \, w^{\mu\nu} = 2 \, e_q^2 \, \delta ( 2q\cdot k -Q^2) \,
(\hat s^2 + \hat u^2) \>. \label{lw}
\ee

In the collinear case, $f_q(x,\kt) = f_q(x) \, \delta^2(\bfk_\perp)$,
Eqs.~(\ref{dsig})--(\ref{lw}) simply result into
\be
\frac{d^2 \sigma ^{\ell p\to \ell X}}{d \xbj \, dQ^2} =
\sum_q f_q(\xbj) \frac{d \hat\sigma^{\ell q\to \ell q}}{dQ^2}\;,
\label{dsigparton}
\ee
where $d\hat\sigma ^{lq\to lq}$ is the cross section for the elementary
lepton-quark scattering,
\be
\frac{d \hat\sigma^{\ell q\to \ell q}}{d Q^2} = e_q^2 \,
\frac{2\pi \alpha^2}{ \xbj^2 s^2}\,
\frac{\hat s^2+\hat u^2}{Q^4}\;\cdot
\label{part-Xsec}
\ee

In the general non collinear case one obtains instead:
\be
\frac{d^2 \sigma ^{\ell p\to \ell X}}{d \xbj \, dQ^2} =
\sum_q \int {d^2 \bfk _\perp}\; f_q(x,\kt) \;
\frac{d\hat\sigma ^{\ell q\to \ell q}}{dQ^2} \;
J(\xbj, Q^2, \bfk_\perp) \>,
\label{dsigkt}
\ee
where $x$ is given in Eq.~(\ref{x}) and the function $J$ is given by
\be
J = \frac{\xbj}{x}
\left( 1 + \frac{\xbj^2}{x^2}\frac{\kt^2}{Q^2} \right)^{\!\!-1} \cdot
\label{fcnj}
\ee
Notice that $J =1$ in the collinear case; the elementary $d\hat\sigma$
in Eq. (\ref{dsigkt}) is the same as in Eq. (\ref{part-Xsec}) with
the Mandelstam variables of Eqs. (\ref{stu}), which depend on $\bfk_\perp$.

If one could detect the final quark -- for example by reconstructing
the current fragmentation jet -- the DIS cross section could be written as
\be
\frac{d^4 \sigma ^{\ell p\to \ell + jet + X}}{d\xbj \, dQ^2 \, d^2\bfk _\perp}
= \sum_q f_q(x,\kt)\;
\frac{d \hat\sigma ^{\ell q\to \ell q}}{dQ^2} \, J(\xbj, Q^2, k_\perp) \>,
\label{dsigjet}
\ee
and one could test the azimuthal dependence of the cross section on the
angle $\varphi$ between the leptonic and the jet plane, Fig. 1. Such a
dependence, resulting from $\hat s^2 + \hat u^2$, Eqs. (\ref{eq:s2}) and
(\ref{eq:u2}), was suggested by Cahn \cite{cahn}, in semi-inclusive DIS,
assuming that the fragmentation process $q \to h X$ is essentially collinear
and the direction of the final detected hadron is close to that of the quark.
Smearing effects due to the fragmentation process were also taken into account
\cite{cahn,kk,aram}. In the next subsection we shall consider SIDIS processes,
fully taking into account the intrinsic motion and the angular dependence
in the fragmentation process; we shall see that indeed a dependence on an
azimuthal angle remains, which allows to obtain an estimate of the average
transverse momentum in distribution and fragmentation functions.

\subsection{SIDIS cross section}

Let us consider semi-inclusive DIS processes, $\ell \, p \to \ell \, h \, X$,
in the $\gamma^* p$ c.m. frame and in the kinematic regime in which 
$P_T \simeq \Lambda_{\rm QCD} \simeq \kt$, where 
$P_T = |\bfP_T|$ is the final hadron transverse momentum. 
In this region leading order elementary processes, 
$\ell \, q \to \ell \, q$, are dominating: the soft $P_T$ of the detected 
hadron is mainly originating from quark intrinsic motion 
\cite{EMC1,EMC2,E665}, rather than from higher order pQCD interactions, 
which, instead, would dominantly produce large $P_T$ hadrons 
\cite{ces,nlo,dfs}. 

We adopt as our starting point a factorized scheme, introducing -- within 
the leading-order parton model with leading-twist distribution and 
fragmentation functions -- the exact $\kt$ kinematics; this induces 
corrections of the ${\cal O}(\kt/Q)$ or higher, which we wish to explore
and compare with experiments. A formal QCD factorization, in the same 
low transverse momentum region and at leading order in $\kt/Q$, has been 
recently proved \cite{ji}.  

\begin{figure}[t]
\begin{center}
\scalebox{0.4}{\input{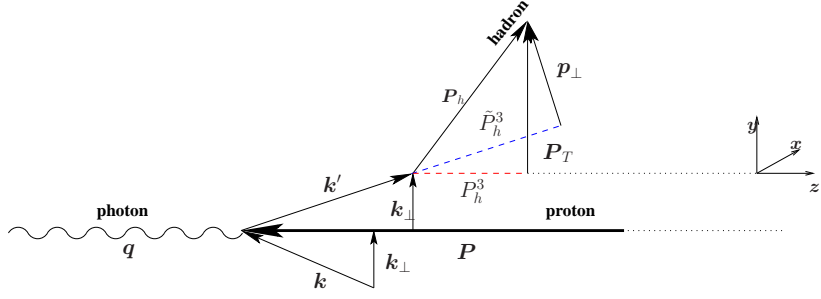}}
\caption{\small Kinematics of the fragmentation process.}
\label{fig:frag}
\end{center}
\end{figure}

\begin{figure}[t]
\begin{center}
\scalebox{0.4}{\input{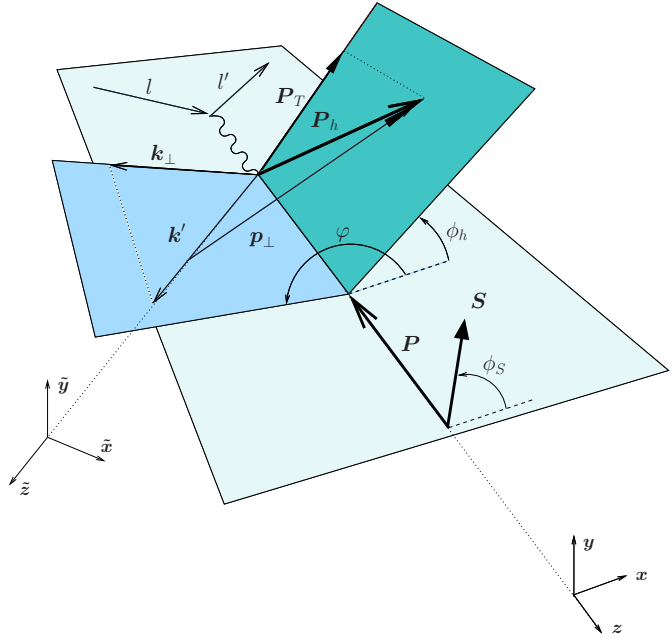}}
\caption{\small Three dimensional kinematics of the SIDIS
process.}
\label{fig:planessidis}
\end{center}
\end{figure}

In the factorization scheme, assuming an independent fragmentation
process, the SIDIS cross section for the production of a hadron $h$ inside
the jet originated from a final quark with transverse momentum $\bfk_\perp$
can be written as
\be
\frac{d^7 \sigma ^{\ell p\to \ell + jet + h + X}}{d\xbj \, dQ^2 \,
d^2 \bfk_\perp
\, dz \, d^2\bfp_\perp} = \sum_q  \> f_q(x,\kt) \;
\frac{d \hat\sigma ^{\ell q\to \ell q}}{dQ^2} \;
J(\xbj, Q^2, k_\perp) \;
D_q^h(z,p _\perp)\,,
\label{sidis-Xsec}
\ee
where $D_q^h(z,\bfp _\perp)$ is the number density of hadrons $h$ resulting
from the fragmentation of the final parton $q$, normalized so that
\be
\int \! dz\,d^2 \bfp _\perp \, D_q^h(z,p _\perp) =
\langle N_h \rangle \,,
\ee
where $\langle N_h \rangle$ is the average multiplicity of hadron $h$ in
the current fragmentation region of quark $q$ and
\be
\int \! d^2 \bfp _\perp \, D_q^h(z,p _\perp) =
D_q^h(z) \,.
\ee
$\bfp _\perp$ is the transverse momentum of the hadron $h$ {\it with respect
to the direction} $\bfk ^\prime$  of the fragmenting quark and $z = P_h^+/k^+$ is the
light-cone fraction of the quark momentum carried by the resulting hadron in the ($\tilde x,\tilde y,\tilde z$)--system (see Figs. \ref{fig:frag} and \ref{fig:planessidis}).
These natural variables for the fragmentation process can be written as:
\bea
z &=& \frac{P_h^0 + \bfP _h \cdot \hat{\bfk}^{\prime}} {2{k^\prime}^0}
\label{z} \\
\bfp _\perp &=&
\bfP _h - (\bfP _h \cdot \hat{\bfk}^\prime) \, \hat{\bfk}^\prime
\label{pt} \,,
\eea
where $P_h = (P_h^0,\bfP_T, P_h^3)$ and
$k' = k + q = ({k^\prime}^0,\bfk _\perp, k^{\prime 3})$, with:
\bea
{k^\prime}^0 &=& \frac{W}{2}
\left(\frac{x-2\xbj+1}{1-\xbj} + \frac{\xbj}{x}\,\frac{\kt^2}{Q^2}\right)
\nonumber \\
k^{\prime 3} &=& \frac{W}{2}
\left(\frac{1-x}{1-\xbj} + \frac{\xbj}{x}\,\frac{\kt^2}{Q^2}\right)
\label{kprime}\\
|\bfk^{\prime}|^2 &=& \kt^2 + (k^{\prime 3})^2 \nonumber \,.
\eea
%
%
\vskip -12pt
The above equations allow us to express, for each value of $\xbj$ and $Q^2$,
the fragmentation variables $z$ and $\bfp_\perp$ in terms of the usual
observed hadronic variables $\bfP_T$ and $z_h = (P\cdot P_h)/(P\cdot q) =
(P_h^0 + P_h^3)/W$. One finds
\bea
z &=& \frac{1}{2{k^\prime}^0} \left[
\left(\frac{z_h \, W}{2} + \frac{P_T^2}{2z_h \, W}\right) +
\frac{\bfP _T \cdot \bfk_\perp}{|\bfk^{\prime}|} +
\frac{k^{\prime 3}}{|\bfk^\prime|}
\left(\frac{z_h \, W}{2} - \frac{P_T^2}{2z_h \, W} \right)
\right] \label{zqh} \\
&=&  z_h  + \frac{\kt P_T}{Q^2} \frac{2\xbj}{1-\xbj}
\, \cos(\phi_h-\varphi) + {\cal O} \left(\frac{\kt^2}{Q^2} \right)
= z_h + {\cal O} \left(\frac{\kt^2}{Q^2} \right)
\label{zqhapp}
\eea
and
\bea
\bfp _\perp &=& \left( \bfP _T -
\frac{\bfP _T \cdot \bfk _\perp + P_h^3 k^{\prime 3}}{|\bfk^{\prime}|^2}\,
\bfk _\perp \,, \;
P_h^3 - \frac{\bfP _T \cdot \bfk _\perp +
P_h^3 k^{\prime 3}}{|\bfk^{\prime}|^2}
\,  k^{\prime 3} \right) \label{ptqh} \\
&=& \bfP _T - z_h \, \bfk _\perp +
{\cal O}\left(\frac{\kt^2}{Q^2} \right) \label{ptqhapp}
\eea
where $k^{\prime 0}, k^{\prime 3}$ and $|\bfk'|$ are given in
Eqs. (\ref{kprime}) and $P_h^3 =(z_h \, W)/2 - P_T^2/(2z_h \, W)$.

%

Eqs. (\ref{zqh}) and (\ref{ptqh}) allow us to describe the fragmentation
process in terms of the variables $(z_h,\bfP _T)$:
\be
dz\;d^2 \bfp _\perp = dz_h \; d^2 \bfP_T \; \frac{z}{z_h} \>,
\ee
so that, finally, the SIDIS cross section (\ref{sidis-Xsec}) can be
written in terms of physical observables as:
\bea
&& \hspace*{-0.8cm}
\frac{d^5\sigma^{\ell p \to \ell h X }}{d\xbj \, dQ^2 \, dz_h \, d^2 \bfP _T} =
\sum_q \int {d^2 \bfk _\perp}\; f_q(x,k_\perp) \;
\frac{d \hat\sigma ^{\ell q\to \ell q}}{dQ^2} \;
J\; \frac{z}{z_h} \; D_q^h(z,p _\perp)
\label{sidis-Xsec-final} \\
&& \hspace*{0.4cm} =
\sum_q  e_q^2 \int  \! d^2 \bfk _\perp \; f_q(x,k _\perp) \;
\frac{2\pi\alpha ^2}{\xbj ^2 s^2}\,\frac{\hat s^2+\hat u^2}{Q^4}\;
D_q^h(z,p _\perp) \; \frac{z}{z_h} \,
\frac{\xbj}{x}\left( 1 + \frac{\xbj^2}{x^2}\frac{\kt^2}{Q^2} \right)^{\!\!-1}
\> \cdot \nonumber
\eea
This is an exact expression at all orders in $(\kt / Q)$;
$x$ is given in Eq. (\ref{x}) and the full expressions of $z$ and $\bfp_T$
in terms of $\xbj, Q^2, \bfk_\perp, z_h$ and $\bfP_T$
can be derived from Eqs. (\ref{kprime}), (\ref{zqh}) and (\ref{ptqh}).
Notice that, in the physical variables $\xbj$ and $z_h$, the $x-z$ 
factorization of Eq. (\ref{sidis-Xsec}) is lost, even in our simple 
parton model treatment; it can be recovered at ${\cal O}(\kt/Q)$ (see 
Eq. (\ref{sidis-Xsec-app}) below). 

Let us now consider again the issue discussed at the end of Section \ref{section-dis},
concerning the azimuthal dependence of the cross section, by comparing
Eqs. (\ref{dsigjet}) and (\ref{sidis-Xsec-final}). The former equation
describes the cross section for jet production and depends, as we explained,
on the azimuthal angle $\varphi$, that is on the azimuthal angle of the
intrinsic $\bfk_\perp$ of the quark in the proton. Such a dependence is
integrated over in Eq. (\ref{sidis-Xsec-final}), which describes the
cross section for the production of a hadron, resulting from the
non collinear fragmentation of the quark. Therefore, there cannot be any
$\varphi$ dependence in this cross section. However, due to relations
(\ref{zqh}) and (\ref{ptqh}), the integration over $\bfk_\perp$ at
fixed $\bfP_T = P_T(\cos\phi_h, \sin\phi_h,0)$ introduces a dependence
on the azimuthal angle $\phi_h$ of the produced hadron $h$, that is
the angle between the leptonic and the hadronic plane, Fig.
\ref{fig:planessidis}. This azimuthal dependence remains in the SIDIS
cross section and will be studied in the next Section (see also Appendix A).

It is instructive, and often quite accurate, to consider the above equations
in the much simpler limit in which only terms of ${\cal O}(\kt/Q)$ are
retained. In such a case $x \simeq \xbj, z \simeq z_h$ and one obtains:
\be
\frac{d^5\sigma^{\ell p \to \ell h X }}{d\xbj \, dQ^2 \, dz_h \, d^2 \bfP _T}
\simeq \label{sidis-Xsec-app}
\sum_q  e_q^2 \int  \! d^2 \bfk _\perp \; f_q(\xbj, k _\perp) \;
\frac{2\pi\alpha ^2}{\xbj ^2 s^2}\,\frac{\hat s^2+\hat u^2}{Q^4}\;
D_q^h(z_h, p _\perp) \> ,
\ee
where $\bfp_\perp \simeq \bfP _T - z_h \, \bfk _\perp$, Eq. (\ref{ptqhapp}),
and
\bea
&& \hat s^2 + \hat u^2 \simeq \xbj ^2 s^2 + (\xbj s + Q^2)^2
- 4 \bfl \cdot \bfk_\perp (2 \xbj s - Q^2) = \nonumber \\
&&\frac{Q^4}{y^2}\left ( 1 + (1-y)^2 -
4 \frac{\kt}{Q}(2-y)\sqrt{1-y}\cos\varphi\right ) \>.
\eea

In what follows we assume, both for the parton densities and the
fragmentation functions, the usual factorization between the
intrinsic transverse momentum and the light-cone fraction
dependences, with a Gaussian $k_\perp$ dependence, that is:
\be
f_q(x,k_\perp) = f_q(x) \, \frac{1}{\pi \langle\kt^2\rangle} \,
e^{-{\kt^2}/{\langle\kt^2\rangle}}
\label{partond}
\ee
and
\be
D_q^h(z,p _\perp) = D_q^h(z) \, \frac{1}{\pi \langle p_\perp^2\rangle}
\, e^{-p_\perp^2/\langle p_\perp^2\rangle}
\label{partonf}
\ee
so that
\be
\int d^2\bfk_\perp \> f_q(x,k_\perp) = f_q(x)
\ee
and
\be
\int d^2\bfp_\perp \> D_q^h(z,p_\perp) = D_q^h(z) \>.
\ee

With the above expressions of $f_q(x,k_\perp)$ and $D_q^h(z,p_\perp)$
the $d^2 \bfk _\perp$ integration in Eq. (\ref{sidis-Xsec-app}) can be
performed analytically, with the result, valid up to ${\cal O}(\kt/Q)$:
\bea
\nonumber &&
\frac{d^5\sigma^{\ell p \to \ell h X }}{d\xbj \, dQ^2 \, dz_h \, d^2\bfP _T}
\simeq
\sum_q \frac{2\pi\alpha^2e_q^2}{Q^4} \> f_q(\xbj) \> D_q^h(z_h) \biggl[
1+(1-y)^2 \\
&& \hskip 36pt - 4 \>
\frac{(2-y)\sqrt{1-y}\> \langle\kt^2\rangle \, z_h \, P_T}
{\langle\pt^2\rangle \, Q}\> \cos \phi_h \biggr]
\frac{1}{\pi\langle\pt^2\rangle} \, e^{-P_T^2/\langle\pt^2\rangle} \, ,
\label{cahn-anal-app}
\eea
where
\be
\langle \pt^2 \rangle = \langle \ptq^2 \rangle + z_h^2 \langle \kt^2 \rangle\,.
\label{eq:meanpt}
\ee

This approximate result illustrates very clearly the origin of the
dependence of the unpolarized SIDIS cross section on the azimuthal angle
$\phi_h$. As observed first by Cahn \cite{cahn,aram}, such a dependence is
related to the parton intrinsic motion and it vanishes when $k_\perp =0$.
Having also taken into account the intrinsic motion in the fragmentation
process, Eq. (\ref{cahn-anal-app}) also depends on $\langle \ptq^2 \rangle$,
via the quantity $\langle \pt^2 \rangle$ defined in Eq. (\ref{eq:meanpt}).

As we said, the above results hold in the small
$P_T \simeq \Lambda_{\rm QCD} \simeq \kt$ region, where corrections
${\cal O}(\kt^2/Q^2)$ are expected to be small. As we shall see in the next
Section the numerical results obtained from Eq. (\ref{sidis-Xsec-final}) or
from Eq. (\ref{cahn-anal-app}) are indeed very close.

\section{Cahn effect in unpolarized SIDIS \label{section-cahn}}

We wish to obtain experimental information on the average
intrinsic motions. Our strategy is that of trying to describe
several sets of experimental data, which explicitly measure the
dependence of the SIDIS unpolarized cross section on the azimuthal
angle $\phi _h$ between the lepton plane and the hadron production
plane, and on the transverse momentum of the detected hadron $P_T$
(see Fig. 3); we do that by exploiting Eq.
(\ref{sidis-Xsec-final}) or (\ref{cahn-anal-app}) and by fixing
the values of $\langle\kt^2\rangle$ and $\langle\ptq^2\rangle$
which best describe the data.

The $\phi_h$ dependence of the SIDIS cross section, for the production of
charged hadrons, has been extensively studied by the EMC collaboration in
the scattering of $280$ GeV muons against a hydrogen target \cite{EMC1,EMC2}.
The shape of the differential cross section
\be
\frac{d\sigma ^{\ell p \to \ell h X}}{d\phi_h}  =
 \int \! d\xbj \, dQ^2 \, dz_h \,  P_T \, d P _T \;
\frac{d^5\sigma^{\ell p \to \ell h X }}{d\xbj \, dQ^2 \, dz_h \, d^2 \bfP _T}
\label{ds/dphi}
\ee
is studied as a function of $\phi_h$.

The integration covers the $\xbj, Q^2, z_h$ and $P_T$ regions
consistent with the experimental cuts
\cite{EMC2}:
\be
\nonumber
x_{F} > 0.1 \quad\quad P_T > 0.2 \; {\rm GeV}/c \quad\quad  y < 0.8
\quad\quad Q^2 > 4 \; ({\rm GeV}/c)^2 \,,
\label{excuts}
\ee
where $x_{F} = {2 P_L}/{W}\,$ and $P_L$ is the longitudinal momentum of
the produced hadron relative to the virtual photon, according to Fig. 3.

In our analysis we adopt the MRST 2001 (LO) \cite{MRST01} parton density
functions and the fragmentation functions into charged hadrons from
Ref. \cite{Kretzer}.
\begin{figure}[t]
\begin{center}
\includegraphics[width=0.5\textwidth]{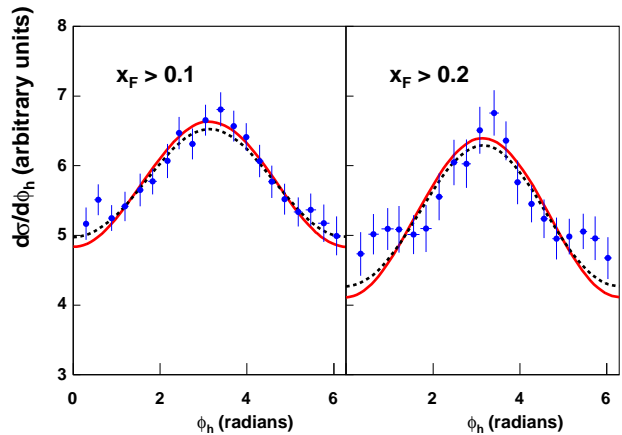}
\caption{\small
Fits to the $\cos\phi_h$ dependence of the cross section: the dashed line
is obtained with exact kinematics, while the solid bold line includes terms
up to ${\cal O}({\kt}/{Q})$ only. The data are from Ref. \cite{EMC2}.}
\label{fig:cahn}
\end{center}
\end{figure}

Fig.~{\ref{fig:cahn}} shows our fits to two sets of data \cite{EMC2}, 
corresponding to two different ranges of $x_F$. The solid bold line shows the  
result we obtain by taking into account only ${\cal O}(\kt/Q)$ contributions, 
Eq. (\ref{cahn-anal-app}), whereas the dashed line corresponds to the exact 
result at all orders in $\kt/Q$, Eq. (\ref{sidis-Xsec-final}). 
The $(-\cos\phi_h)$ behaviour,
explicit in Eq. (\ref{cahn-anal-app}), is indeed shown by the data.
A possible positive contribution from a $\cos(2\phi_h)$ term seems to be 
visible at small values of $\phi_h$ (dashed line). 

\begin{figure}[t]
\begin{center}
\includegraphics[width=0.7\textwidth,bb= 10 400 540 660]{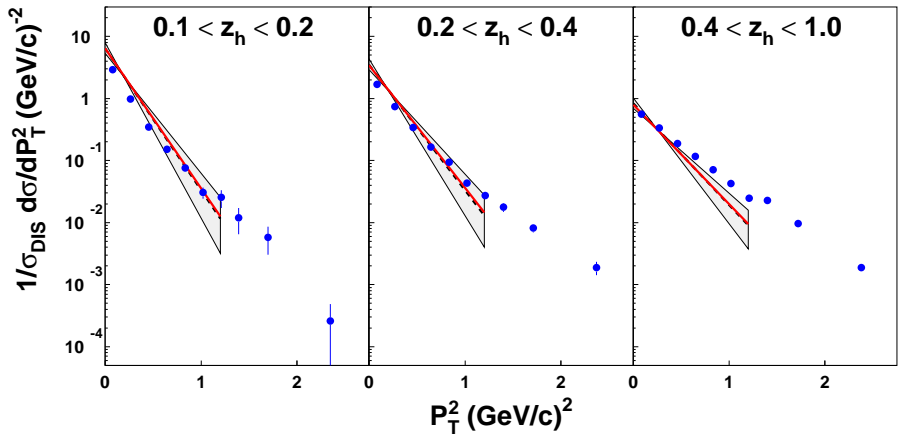}
\caption{\small
The normalized cross section $d\sigma/dP_T^2$: the dashed line
is obtained with exact kinematics, while the solid bold line includes terms
up to ${\cal O}({\kt}/{Q})$ only. The shadowed region corresponds to varying
the parameters $\langle\kt^2\rangle$ and $\langle\ptq^2\rangle$, given in
Eq. (\ref{parameters}) of the text, by 20\%. The data are from Ref. \cite{EMCpt}.}
\label{fig:pt}
\end{center}
\end{figure}

Other interesting EMC data \cite{EMCpt} concern the $P_T^2$ dependence of
the cross section for $\mu p$ and $\mu d$ scattering at incident beam energy
between $100$ and $280$ GeV; also these data are strictly related to the
values of $\langle\kt^2\rangle$ and $\langle\ptq^2\rangle$.
The quantity measured is given by
\be
\frac{1}{\sigma_{DIS}} \, \frac{d\sigma}{dP_T^2} = \frac{1}{2\,\sigma_{DIS}}
\, \int \! d\phi_h \, d\xbj \, dQ^2 \, dz_h \;
\frac{d^5\sigma^{\ell p \to \ell h X }}{d\xbj \, dQ^2 \, dz_h \, d^2 \bfP _T},
\label{eq:sigmapt}
\ee
where $\sigma_{DIS}$ is the integrated DIS cross section
from Eq. (\ref{dsig}).
In the integration of
Eq. (\ref{eq:sigmapt}) the following experimental cuts have been imposed
(see Ref.~\cite{EMCpt} for further details):
\bea
\nonumber
Q^2 > 5 \; ({\rm GeV}/c)^2 \;,\;\;\; W^2 < 90 \; {\rm GeV^2}\;,\;\;\;
E_h > 5 \; {\rm GeV} \\
0.1 < z_h < 0.9 \;,\;\;\; 0.2 < y < 0.8 \>.
\eea

Fig.~\ref{fig:pt} shows the comparison of our results with EMC data 
\cite{EMCpt} for different ranges of $z_h$. The solid and dashed lines,
which are here basically indistinguishable, are the results of our fits
at first order and at all orders in $\kt/Q$ respectively.
The shadowed region is spanned by varying the parameters 
$\langle\kt^2\rangle$ and $\langle\ptq^2\rangle$ by $20$\% and shows the 
sensitivity of our results on these parameters. 
The figure clearly show
that, as expected, our LO approach is valid for $P_T$ values up to about
$1$ GeV/$c$. At higher values NLO contributions from $\gamma^* q \to g q$
and $\gamma^* g \to q \bar q$ processes have to be taken into account.

Useful data on the $\phi_h$ and $P_T$ dependence were also found by the
FNAL E665 collaboration \cite{E665} in $\mu p$ and $\mu d$ interactions
at $490$ GeV. The quantity studied is
\be
\langle\cos\phi_h\rangle =
\frac{\int  d\xbj \, dQ^2 \, dz_h \, d^2\bfP_T \, \cos\phi_h \, d^5\sigma}
{\int  d\xbj \, dQ^2 \, dz_h \, d^2\bfP_T \, d^5\sigma}
\label{eq:cosphi}
\ee
where $d^5\sigma$ denotes the fully differential cross section
\be
d^5\sigma \equiv
\frac{d^5\sigma^{\ell p \to \ell h X }}{d\xbj \, dQ^2 \, dz_h \, d^2\bfP _T}
\> , \label{d5s}
\ee
and where the integral over $P_T$ runs from $P_T^{cut}$ to
$P_T^{max}\simeq 2.5$~GeV/$c$. According to the experimental setup
\cite{E665}, the integration region in Eq. (\ref{eq:cosphi}) is defined by:
\bea
\nonumber
Q^2 > 3 \; ({\rm GeV}/c)^2 \;,\;\;\; 300 < W^2 < 900 \; {\rm GeV^2}\;,\\
60 < \nu < 500 \; {\rm GeV}\;,\;\;\;
E_h > 8 \; {\rm GeV} \;,\;\;\; 0.1 < y < 0.85 \>.
\eea

\begin{figure}[t]
\begin{center}
\includegraphics[width=0.47\textwidth]{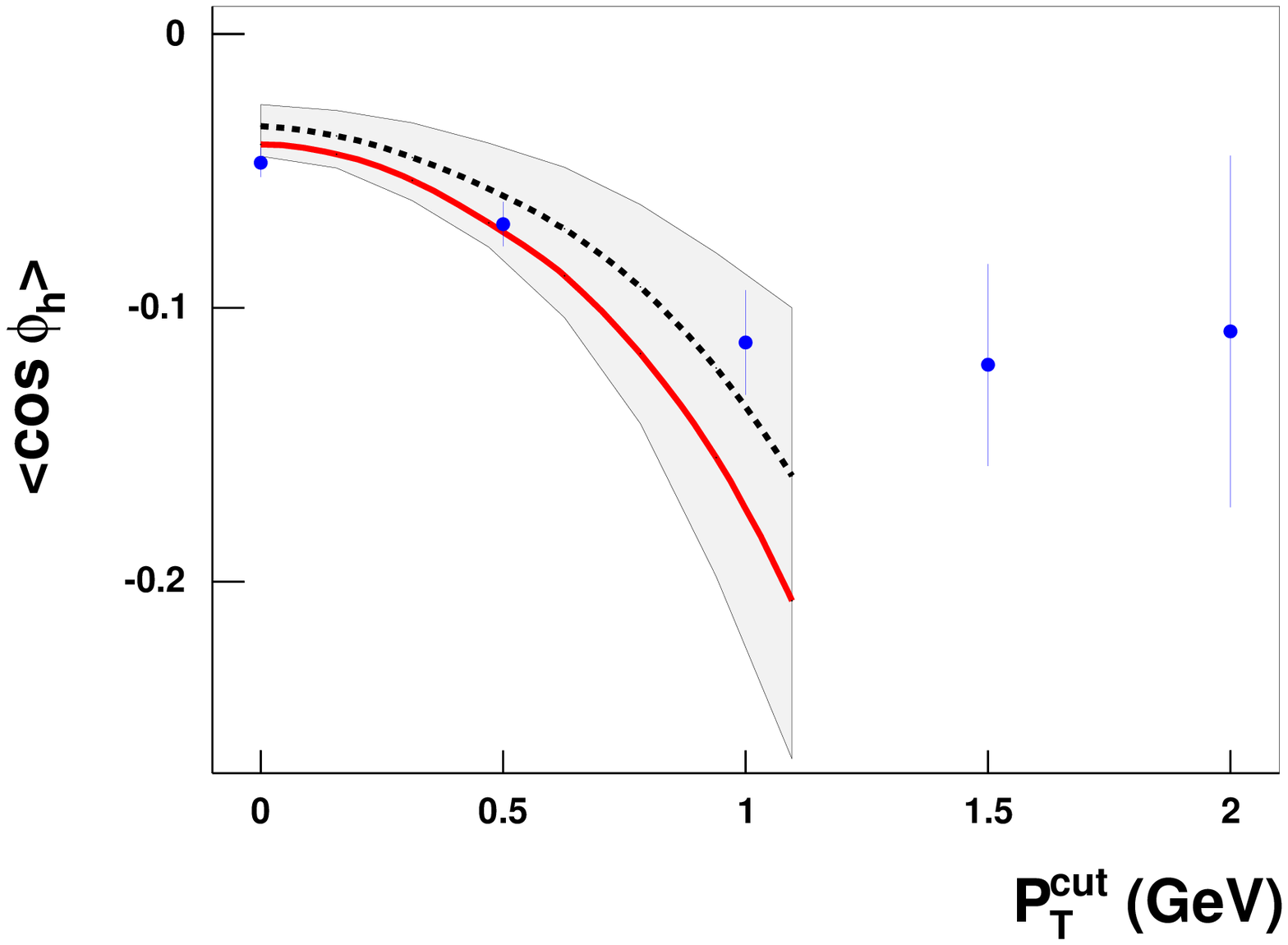}
\hfill~\includegraphics[width=0.47\textwidth]{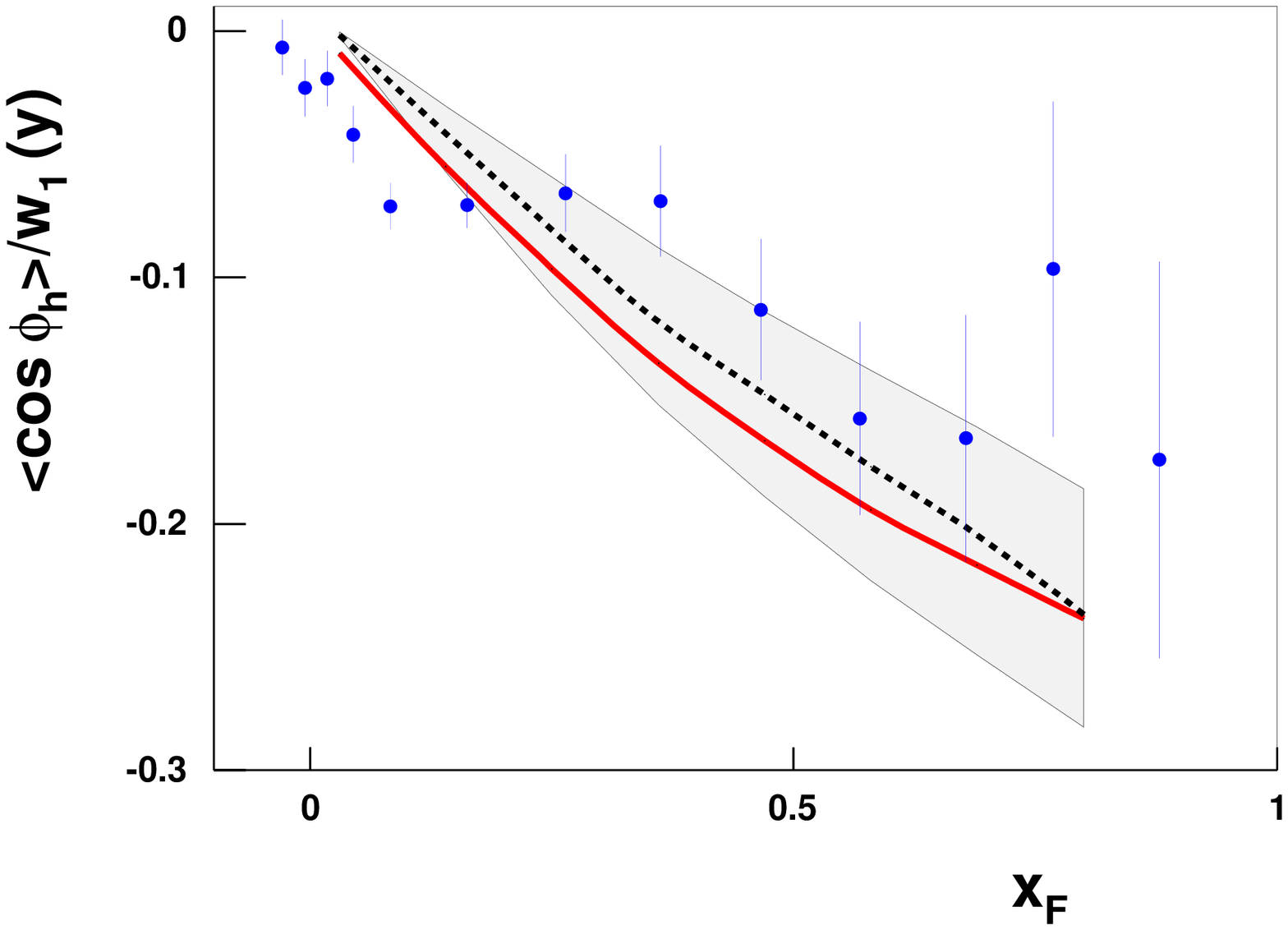}
\caption{\small
$\langle \cos\phi_h \rangle$, as given by Eq. (\ref{eq:cosphi}),
as a function of $P_T^{cut}$, left panel, and $x_F$, right panel: the dashed 
line
is obtained with exact kinematics, while the solid bold line includes terms
up to ${\cal O}({\kt}/{Q})$ only. The shadowed region corresponds to varying
the parameters $\langle\kt^2\rangle$ and $\langle\ptq^2\rangle$, given in
Eq. (\ref{parameters}) of the text, by 20\%. 
$w_1(y) = \frac{(2-y)\sqrt{1-y}}{1+(1-y)^2}$. The data are from Ref. \cite{E665}, left panel,
and Ref. \cite{EMC2}, right panel.}
\label{fig:cosphi}
\end{center}
\end{figure}

Fig.~\ref{fig:cosphi} shows 
the data \cite{E665} as a function of $P_T^{cut}$ compared with the results of our
calculations; again, the solid bold 
line corresponds to
the result we obtain by taking into account only $O(\kt/Q)$ terms,
Eq. (\ref{cahn-anal-app}), whereas the dashed line corresponds to the exact
kinematics, Eq. (\ref{sidis-Xsec-final}). 
We also show the EMC data on the $x_F$ dependence of
$\langle \cos\phi_h \rangle/ w_1 (y)$ \cite{EMC2}; they compare well with our results obtained by using Eq. (\ref{eq:cosphi}) (in the right panel of Fig.~\ref{fig:cosphi} the theoretical curve corresponds to the calculation of $\langle \cos\phi_h \rangle/ \langle w_1 (y) \rangle$) with the experimental cuts of Ref. \cite{EMC2}  and
no integration over $z_h$ which is expressed in terms of $x_F$. The shadowed region
is obtained by varying the parameters $\langle\kt^2\rangle$ and
$\langle\ptq^2\rangle$ by $20$\%. Once more, Fig.~\ref{fig:cosphi} clearly
shows that our calculation is valid for $P_T$ values up to about $1$ GeV/$c$,
where NLO pQCD contributions \cite{ces, nlo} must be taken into account and
our simple LO treatment can no longer be applied.

All sets of data described above depend crucially on the intrinsic motion
in distribution and fragmentation functions; their combined analysis
leads to the following best values of the parameters:
\be
\langle\kt^2\rangle   = 0.25  \;({\rm GeV}/c)^2 \quad\quad\quad
\langle\ptq^2\rangle  = 0.20 \;({\rm GeV}/c)^2 \>.
\label{parameters}
\ee

One should notice that the above values have been derived from
sets of data collected at different energy, $\xbj$, $Q^2$ and
$z_h$ ranges, looking at the combined production of all charged
hadrons in SIDIS processes, and assuming constant values of
$\langle\kt^2\rangle$ and $\langle\ptq^2\rangle$, which allow
analytical integration up to ${\cal O}(\kt/Q)$; these values are
also assumed to be independent of the (light) quark flavor. A more
refined analysis, introducing for example $x$ and $z$
dependences, would require the introduction of new unknown
functions. At this stage, we stick to the constant best values
(\ref{parameters}) which, together with Eqs. (\ref{partond}) and
(\ref{partonf}), can only be considered as a consistent simple
estimate and a convenient parametrization of the true intrinsic
motion of quarks in nucleons and of hadrons in jets, supported by
the available experimental information. In the next Section we
adopt such a picture for the computation of pion and kaon SSA in
SIDIS polarized processes, $\ell \, \pup \to \ell \, \pi \, X$ and
$\ell \, \pup \to \ell \, K \, X$.

\section{Sivers effect in polarized SIDIS}

In this Section we consider the single spin asymmetry $A_{UT}$ measured
in $e \pup \to e \pi X$ processes by the HERMES collaboration at DESY, using
a transversely polarized proton target \cite{hermUT}. Our aim is that of
obtaining information on the quark Sivers functions, which can be isolated
and directly accessed by studying the weighted transverse spin asymmetry
$A_{UT}^{\sin(\phi_\pi - \phi_S)}$ \cite{bm}. These functions are then
compared with those obtained from the study of SSA in $\pup p$ processes. They
are also used to estimate the contribution of the Sivers mechanism alone to
the weighted SSA $A_{UL}^{\sin\phi_\pi}$, measured by the HERMES collaboration
in the lepton scattering off a longitudinally polarized proton target
\cite{hermUL}. Moreover, expectations for $A_{UT}^{\sin(\phi_\pi - \phi_S)}$
in the COMPASS kinematical regions are computed and compared with preliminary
COMPASS data \cite{compUT} and predictions for kaon asymmetries
$A_{UT}^{\sin(\phi_K - \phi_S)}$ at HERMES are given.

Let us recall the origin of the SSA in SIDIS, as originated by the Sivers
mechanism \cite{siv}. The unpolarized quark (and gluon) distributions
inside a transversely polarized proton (generically denoted by $\pup$, with
$\pdown$ denoting the opposite polarization state) can be written as:
\be
f_ {q/\pup} (x,\bfk_\perp) = f_ {q/p} (x,\kt) +
\frac{1}{2} \, \Delta^N f_ {q/\pup}(x,\kt)  \;
{\bfS} \cdot (\hat {\bfP}  \times
\hat{\bfk}_\perp)\; , \label{poldf}
\ee
where $\bfP$ and $\bfS$ are respectively the proton momentum and transverse
polarization vector, and $\bfk_\perp$ is the parton transverse momentum;
transverse refers to the proton direction. Eq. (\ref{poldf}) implies
\bea
\nonumber
&&f_ {q/\pup} (x,\bfk_\perp) + f_ {q/\pdown} (x,\bfk_\perp) =
2 f_ {q/p} (x,\kt)\;, \\
&&f_ {q/\pup} (x,\bfk_\perp) - f_ {q/\pdown} (x,\bfk_\perp) =
\Delta^N f_ {q/\pup}(x,\kt)\;{\bfS} \cdot (\hat{\bfP}  \times
\hat {\bfk}_\perp)\;, \label{sivf}
\eea
where $f_ {q/p} (x,\kt)$ is the unpolarized parton density and
$\Delta^N f_ {q/\pup}(x,\kt)$ is referred to as the Sivers function.
Notice that, as requested by parity invariance, the scalar quantity
$\bfS \cdot (\hat{\bfP}  \times \hat {\bfk}_\perp)$ singles out the
polarization component perpendicular to the $\bfP-\bfk_\perp$ plane.
For a proton moving along $-z$ and a generic transverse polarization
vector $\bfS = |\bfS|\,(\cos\phi_S, \sin\phi_S, 0)$
(see Fig. \ref{fig:planessidis}) one has:
\be
\bfS \cdot (\hat{\bfP}  \times \hat{\bfk}_\perp) = |\bfS| \,
\sin(\varphi-\phi_S) \equiv |\bfS| \, \sin\phi_{Siv} \,,
\ee
where $(\varphi-\phi_S) = \phi_{Siv}$ is the Sivers angle.

The cross section for the scattering of an unpolarized lepton off a polarized
proton, in the configuration of Fig. \ref{fig:planessidis}, can then simply
be written as, see Eq. (\ref{sidis-Xsec-final}),
\be
d^6\sigma^{\uparrow} = \frac{1}{2\pi}
\sum_q \int {d^2 \bfk _\perp}\;f_{q/\pup}(x,k _\perp) \;
\frac{d \hat\sigma ^{\ell q\to \ell q}}{dQ^2} \;
J\; \frac{z}{z_h} \; D_q^h(z,p_\perp) \label{polcs}
\ee
where $d^6\sigma^{\uparrow}$ stands for
\be
d^6\sigma^{\uparrow} \equiv
\frac{d^6\sigma^{\ell \pup \to \ell h X }}
{d\xbj \, dQ^2 \, dz_h \, d^2\bfP _T \, d\phi_S}
\> , \label{d5ps}
\ee
and where the $\phi_S$ dependence is contained in $f_{q/\pup}(x,\bfk _\perp)$.
The $\phi_S$ dependence originates from the fact that, with transversely
polarized protons, the cross section depends also on the angle between the
polarization vector and the leptonic plane; in the configuration of
Fig. 3 this is simply $\phi_S - \phi_{\ell'}$ = $\phi_S$, having chosen
$\phi_{\ell'}$ to be zero, with $\phi_S$ varying event by event. In actual
experiments variables are measured in a different frame (for example the
laboratory frame where the lepton moves along the $z$-axis and the proton
is at rest); a comprehensive set of relations between the different frames
can be found in Ref. \cite{aram}. Let us only notice here that the difference
between the azimuthal angles of the final lepton in the frame of Fig. 3 and
in the laboratory frame is of the ${\cal O}(\kt^2/Q^2)$.

This leads to the possibility of a non vanishing transverse single
spin asymmetry, the analyzing power
\be
A = \frac{d^6\sigma ^{\uparrow} - d^6\sigma^{\downarrow}}
           {d^6\sigma ^{\uparrow} + d^6\sigma^{\downarrow}} \>,
\ee
which is given, according to Eqs. (\ref{polcs}) and (\ref{sivf}) by:
\be
A =
\frac{\displaystyle \sum_q \int {d^2 \bfk _\perp}\;
\Delta ^N f_{q/\pup} (x,\kt) \;
[\bfS \cdot (\hat {\bfP} \times \hat{\bfk}_\perp)] \,
\frac{d \hat\sigma ^{\ell q\to \ell q}}{dQ^2} \; J\;
\frac{z}{z_h} \; D_q^h(z,p_\perp) }
{\displaystyle 2 \sum_q \int {d^2 \bfk _\perp}\; f_q(x,\kt) \;
\frac{d \hat\sigma ^{\ell q\to \ell q}}{dQ^2} \;
J\; \frac{z}{z_h} \; D_q^h(z,p_\perp) } \> \cdot \label{anp}
\ee

The above equation gives the transverse SSA originated by the
Sivers mechanism alone, properly taking into account all intrinsic motions.
As it is written it depends on $\xbj, Q^2, z_h, \bfp_{_T}$ and $\phi_S$: of
course, in order to increase statistics and according to the experimental
setups, both the numerator and denominator of Eq. (\ref{anp}) can be
integrated over some of the variables.

HERMES first data \cite{hermUL} were gathered in the scattering of unpolarized
leptons ($U$) off ``longitudinally'' ($L$) polarized proton, where
``longitudinal'' means {\it antiparallel} to the lepton direction in
the proton rest frame. Such a direction has a small transverse component, in
the $\gamma^* p$ frame, with respect to the proton direction, that is
\be
\bfS = - \sin\theta_\gamma \, (1,0,0)   \quad\>
\bfS \cdot (\hat {\bfP} \times \hat{\bfk}_\perp) =
- \sin\theta_\gamma \, \sin\varphi \quad\>
\sin\theta_\gamma \simeq \frac{2 \, m_p \, \xbj}{Q} \, \sqrt{1-y} \>.
\label{sul}
\ee

HERMES data are presented for the $\sin \phi_h$ moment of the
analyzing power,
\be
A^{\sin\phi_h} = 2 \,
\frac{\int d\phi_h \, [d\sigma ^{\uparrow} - d\sigma^{\downarrow}]
      \, \sin\phi_h}
     {\int d\phi_h \, [d\sigma ^{\uparrow} + d\sigma^{\downarrow}]} \>,
\label{sina}
\ee
which, from Eq. (\ref{anp}), is
\be
A^{\sin\phi_h}_{UL} =
\frac{\displaystyle - \sum_q \int {d\phi_h \, d^2 \bfk _\perp}\;
\Delta ^N f_{q/\pup} (x,\kt) \sin \theta _\gamma \sin \varphi \;
\frac{d \hat\sigma ^{\ell q\to \ell q}}{dQ^2} \;
J\; \frac{z}{z_h} \; D_q^h(z,p_\perp) \sin \phi_h }
{\displaystyle  \sum_q \int {d\phi_h \,d^2 \bfk _\perp}\; f_q(x,k _\perp) \;
\frac{d \hat\sigma ^{\ell q\to \ell q}}{dQ^2} \;
J\; \frac{z}{z_h} \; D_q^h(z,p_\perp) }
\label{hermesul}
\ee
where both numerator and denominator can be integrated over some of the
variables. Eq. (\ref{hermesul}) gives the contribution of the Sivers function
to $A^{\sin\phi_h}_{UL}$. Notice that it is kinematically suppressed by
the $\sin \theta _\gamma$ value, Eq. (\ref{sul}), and that other contributions
might be equally important; they can originate from the Collins mechanism or
from higher-twist terms.

More recently, data were obtained with a transversely polarized
($T$) proton target \cite{hermUT}, $\bfS \cdot (\hat {\bfP} \times
\hat{\bfk}_\perp) = \sin(\varphi - \phi_S)$, and presented
for the $\sin(\phi_h - \phi_S)$ moment of the analyzing power,
which singles out the Sivers contribution \cite{bm}. Eq.
(\ref{anp}) in this case gives
\bea
&& \!\!\!\! A^{\sin (\phi_h-\phi_S)}_{UT} = \label{hermesut} \\
&& \hskip-18pt \frac{\displaystyle  \sum_q \int
{d\phi_S \, d\phi_h \, d^2 \bfk _\perp}\;
\Delta ^N f_{q/\pup} (x,\kt) \sin (\varphi -\phi_S) \;
\frac{d \hat\sigma ^{\ell q\to \ell q}}{dQ^2} \;
J\; \frac{z}{z_h} \; D_q^h(z,p_\perp) \sin (\phi_h -\phi_S) }
{\displaystyle \sum_q \int {d\phi_S \,d\phi_h \, d^2 \bfk _\perp}\;
f_q(x,k _\perp) \; \frac{d \hat\sigma ^{\ell q\to \ell q}}{dQ^2} \;
J\; \frac{z}{z_h} \; D_q^h(z,p_\perp) } \> \cdot \nonumber
\eea

We use Eq. (\ref{hermesut}) to compute $A^{\sin(\phi_h - \phi_S)}_{UT}$,
which can only receive contributions from the Sivers mechanism, and
compare it with data, in order to gather information on the Sivers function
$\Delta ^N f_{q/\pup} (x,\kt)$.

\subsection{Parameterization of the Sivers function}

We parameterize, for each light quark flavour $q=u,d$, the
Sivers function in the following factorized form \cite{col,noiD}:
\be
\Delta^N f_ {q/\pup}(x,\kt) = 2 \, {\cal N}_q(x) \, h(\kt) \,
f_ {q/p} (x,\kt)\; , \label{sivfac}
\ee
where
\bea
&&{\cal N}_q(x) =  N_q \, x^{a_q}(1-x)^{b_q} \,
\frac{(a_q+b_q)^{(a_q+b_q)}}{a_q^{a_q} b_q^{b_q}}\; ,
\label{siversx} \\
&&h(\kt) = \frac{2\kt \, M}{\kt^2+ M^2}\;  ,
\label{siverskt}
\eea
where $N_q$, $a_q$, $b_q$ and $M$ (GeV/$c$) are parameters. $f_ {q/p} (x,\kt)$
is the unpolarized distribution function defined in Eq.~(\ref{partond}).
Since $h(\kt) \le 1$ and since we allow the constant parameter $N_q$ to
vary only inside the range $[-1,1]$ so that $|{\cal N}_q(x)| \le 1$ for any
$x$, the positivity bound for the Sivers function is automatically fulfilled:
\be
\frac{|\Delta^N f_ {q/\pup}(x,\kt)|}{2 f_ {q/p} (x,\kt)}\le 1\; .
\ee

\begin{figure}[H]
\begin{center}
\includegraphics[width=0.5\textwidth, height=0.3\textheight, 
bb= 10 140 540 660]{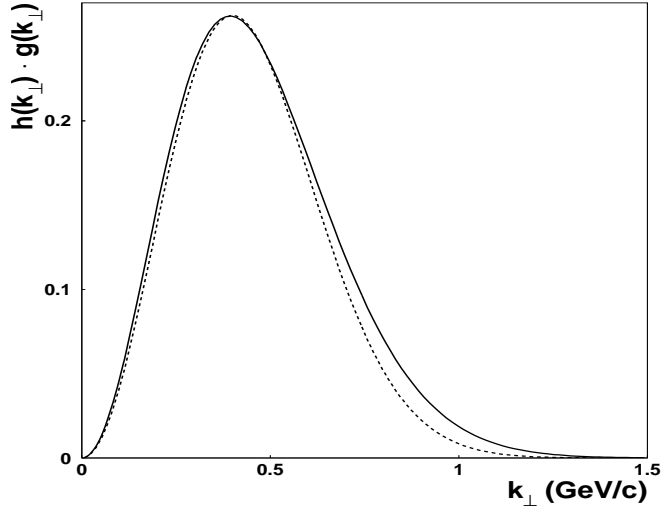}
\caption{\small The product $h(\kt)\cdot g(\kt)$, solid line,
compared with the product $h'(\kt)\cdot g(\kt)$, dashed line, as a
function of $\kt$.} \label{fig:param}
\end{center}
\end{figure}

As an alternative parameterization for $h(\kt)$ we have also used
\be
h'(\kt) = \sqrt{2e} \, \frac{\kt} {M'} \, e^{-\kt^2/M'^{2}}.
\label{siverskt1}
\ee
The two parameterizations (\ref{siverskt}) and (\ref{siverskt1}) are
indeed equivalent at low $\kt$, but they differ at large values
of $\kt$. Nevertheless, we have checked that, once multiplied by the Gaussian
function of $\kt$ contained in the definition of $f(x,\kt)$ [see
Eq.~(\ref{partond})], they give basically the same $\kt$ dependence to the
Sivers function over the whole range, as shown in Fig. \ref{fig:param}
($M^2 = 0.25$ (GeV/$c$)$^2$ and $M^{\prime 2} = 0.36$ (GeV/$c$)$^2$).

The parameterization of Eq. (\ref{siverskt1}) allows to  easily
perform, at ${\cal O}(\kt/Q)$, an analytical integration of Eq.
(\ref{hermesut}), leading to an explicit approximate expression
for the single spin asymmetry:
\be
A_{UT}^{\sin (\phi_h-\phi_S)}(\xbj, z_h, P_T) \simeq
\frac{\Delta\sigma_{Siv}}{\sigma_{0}} \; ,
\label{siversutapp}
\ee
where
\bea
\nonumber
\displaystyle \Delta\sigma_{Siv}(\xbj, y, z_h, P_T)  =
\frac{2 \pi \alpha^2}{\xbj \, y^2 s} \,
 \left( 1+(1-y)^2 \right) \sum_q e_q^2 \, 2 {\cal N}_q(\xbj) \, f_q(\xbj) \,
D_q^h(z_h) \\
\displaystyle  z_h P_T \frac{\sqrt{2e} \widehat{\langle \kt ^2 \rangle}^2 }
{ M' \widehat{ \langle P_T ^2 \rangle}^2 \langle \kt ^2 \rangle}
\exp{\left(-\frac{P_T^2}{\widehat{\langle P_T ^2 \rangle}}\right)} ,
\label{sivers1app}
\eea
and
\bea
\nonumber
\displaystyle \sigma_{0}(\xbj, y, z_h, P_T)  = 2 \pi \frac{2 \pi \alpha^2}
{\xbj \, y^2 s} \,  \left( 1+(1-y)^2 \right) \sum_q e_q^2 \, f_q(\xbj) \,
D_q^h(z_h) \\
\displaystyle \frac{1}{
\pi \langle P_T ^2 \rangle}
\exp{\left(-\frac{P_T^2}{\langle P_T ^2 \rangle}\right)} ,
\label{sivers0app}
\eea
where
\be
\widehat{\langle \kt^2 \rangle} = \frac{M^{\prime 2} \, \langle \kt^2 \rangle}
{M^{\prime 2} + \langle \kt ^2 \rangle} ,\quad
\widehat{\langle P_T^2 \rangle} = \langle \ptq^2 \rangle +
z_h^2 \, \widehat{\langle \kt ^2 \rangle} \,, \label{parsiversutapp}
\ee
and $\langle P_T^2 \rangle$ is given in Eq. (\ref{eq:meanpt}).
Eq. (\ref{sivers1app}) shows that $A_{UT}^{\sin (\phi_h-\phi_S)} = 0$
when $z_h = 0$ or $P_T = 0$.

\subsection{HERMES data and Sivers functions}

Let us now try to understand the HERMES data on
$A^{\sin (\phi_\pi-\phi_S)}_{UT}$ \cite{hermUT}, according to
Eq. (\ref{hermesut}) (exact kinematics) or Eqs.
(\ref{siversutapp})--(\ref{parsiversutapp}) (kinematics up to
${\cal O}(\kt/Q)$). As in Section \ref{section-cahn}, the unpolarized distribution functions
are taken from Ref. \cite{MRST01} and the fragmentation functions from
Ref. \cite{Kretzer}. In the numerator we take into account only the Sivers
contribution of $u$ and $d$ quarks and antiquarks, with separate valence and 
sea functions.
More precisely, we adopt the following form for the Sivers functions:
\be
\Delta^N f_ {q/\pup}(x,\kt) = 2 \, {\cal N}_q(x) \, h'(\kt) \,
f_ {q/p} (x,\kt)\; , \label{sivfunc}
\ee
where ${\cal N}_q$ is given in Eq. (\ref{siversx}), $h'$ in
Eq. (\ref{siverskt1}) and $q = u_v, d_v, u_s, d_s, \bar u, \bar d$.
For the sea quark contributions we assume:
\be
\Delta^N f_{q_s/\pup}(x,\kt) = \Delta^N f_ {\bar q/\pup}(x,\kt)\;,
\label{sivsea}
\ee
for a total of 4 unknown functions, each depending on 3 parameters;
in addition, $h'(\kt)$ depends on the parameter $M'$.

We fit the HERMES data on $A^{\sin (\phi_\pi-\phi_S)}_{UT}$ exploiting the
simplified expressions (\ref{siversutapp})--(\ref{parsiversutapp}).
The resulting best values for the 13 free parameters are shown in Table 1.
\begin{table}[t]
\begin{center}
\begin{tabular}{|ll|ll|}
\hline
$N_{u_v}$ = & $0.42  \pm  0.18$ & $N_{d_v}$ = & $-1.0  \pm  1.8 $ \\
$a_{u_v}$ = & $0.0 \pm   3.3$ & $a_{d_v}$ = & $1.1 \pm  1.2$ \\
$b_{u_v}$ = & $2.6  \pm  1.8$ & $b_{d_v}$ = & $5.0 \pm  3.6$ \\
\hline
$N_{\bar u}$ = & $1.0  \pm  1.9$ & $N_{\bar d}$ = & $-1.0  \pm  1.9 $ \\
$a_{\bar u}$ = & $0.52   \pm  0.43  $ & $a_{\bar d}$ = & $0.0 \pm  4.5$ \\
$b_{\bar u}$ = & $0.0   \pm  3.1$ & $b_{\bar d}$ = & $0.0 \pm  2.8$  \\
\hline
$M'^2$ = & $0.36 \pm 0.43 \; {\rm (GeV}/c)^2$ & $\chi^2/{d.o.f.}$ = & $0.89$  \\
\hline
\end{tabular}
\end{center}
\caption{Best values of the parameters of the Sivers functions.
\label{fitpar}}
\end{table}
The errors are generated by the MINUIT minimizer.
The large errors reflect the large errors of the data and the scarce available
information.

\begin{figure}[T]
\begin{center}
\includegraphics[width=1.\textwidth,bb= 10 140 540 660]
{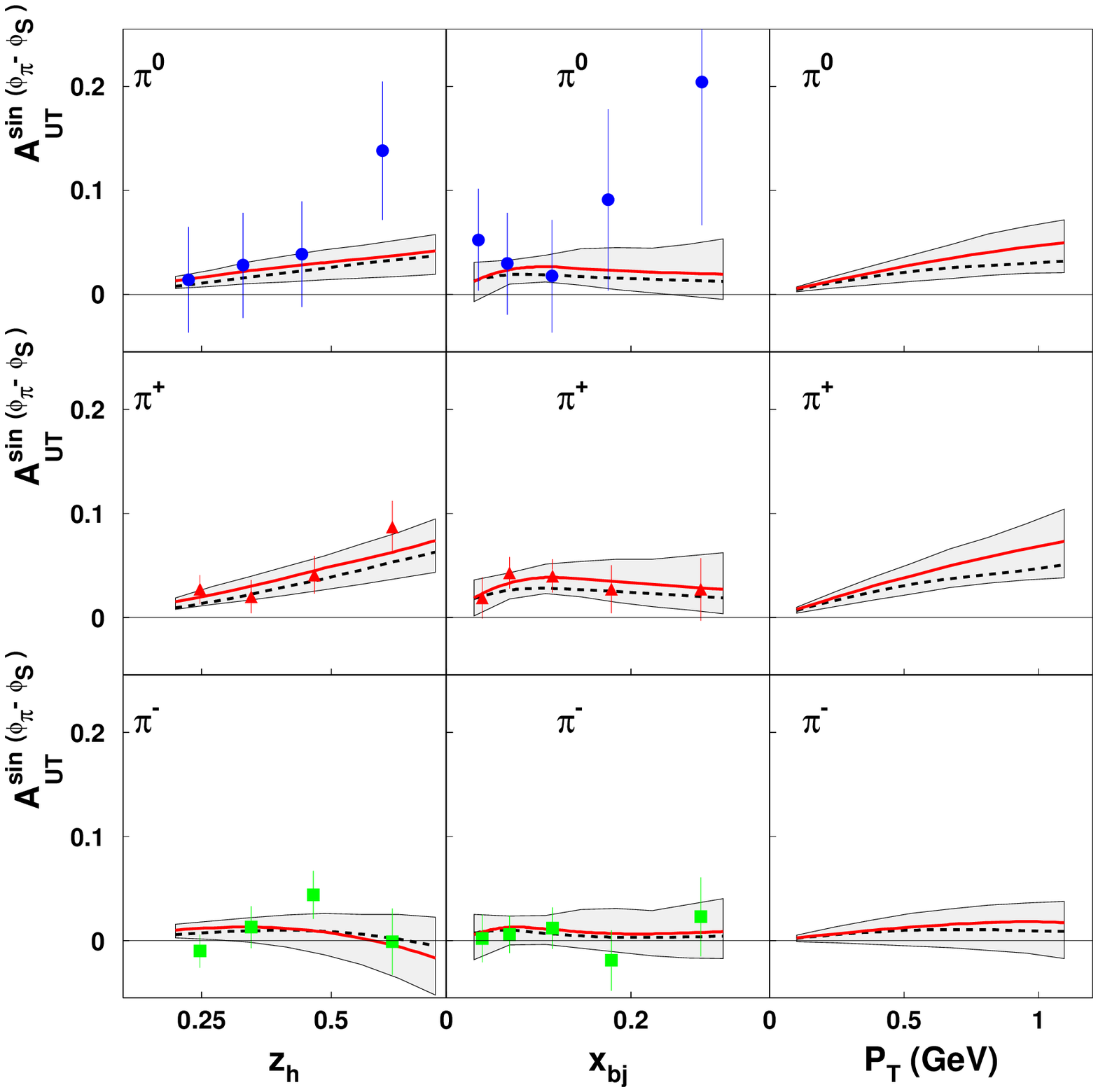}
\caption{\small
HERMES data on $A_{UT}^{\sin(\phi_\pi-\phi_S)}$ \cite{hermUT} for scattering
off a transversely polarized proton target and pion production. The curves are
the results of our fits, with exact kinematics (dashed line) or keeping only
terms up to ${\cal O}({\kt}/{Q})$ (solid bold line). The shadowed region
corresponds to the theoretical uncertainty due to the parameter errors. Data
on the $P_T$ dependence of $A_{UT}^{\sin(\phi_\pi-\phi_S)}$ are not available
yet, and the third column of the figure gives our predictions.}
\label{fig:authermes}
\end{center}
\end{figure}

Our fit is shown in Fig. \ref{fig:authermes}. The solid bold line takes
into account terms up to ${\cal O}(\kt/Q)$, the dashed line is obtained
with the full exact $\bfk_\perp$ kinematics, Eq. (\ref{hermesut}).
In both cases the parameters of Table 1 are used. The shadowed region
corresponds to one-sigma deviation at 90\% CL and was calculated using the
errors (Table \ref{fitpar}) and the parameter correlation matrix generated
by MINUIT, minimizing and maximizing the function under consideration, in a
13-dimensional parameter space hyper-volume corresponding to one-sigma
deviation. Notice that, as expected, the results obtained with exact or
approximate kinematics are very similar.

\begin{figure}[T]
\begin{center}
\includegraphics[width=1.\textwidth,bb= 10 140 540 660]
{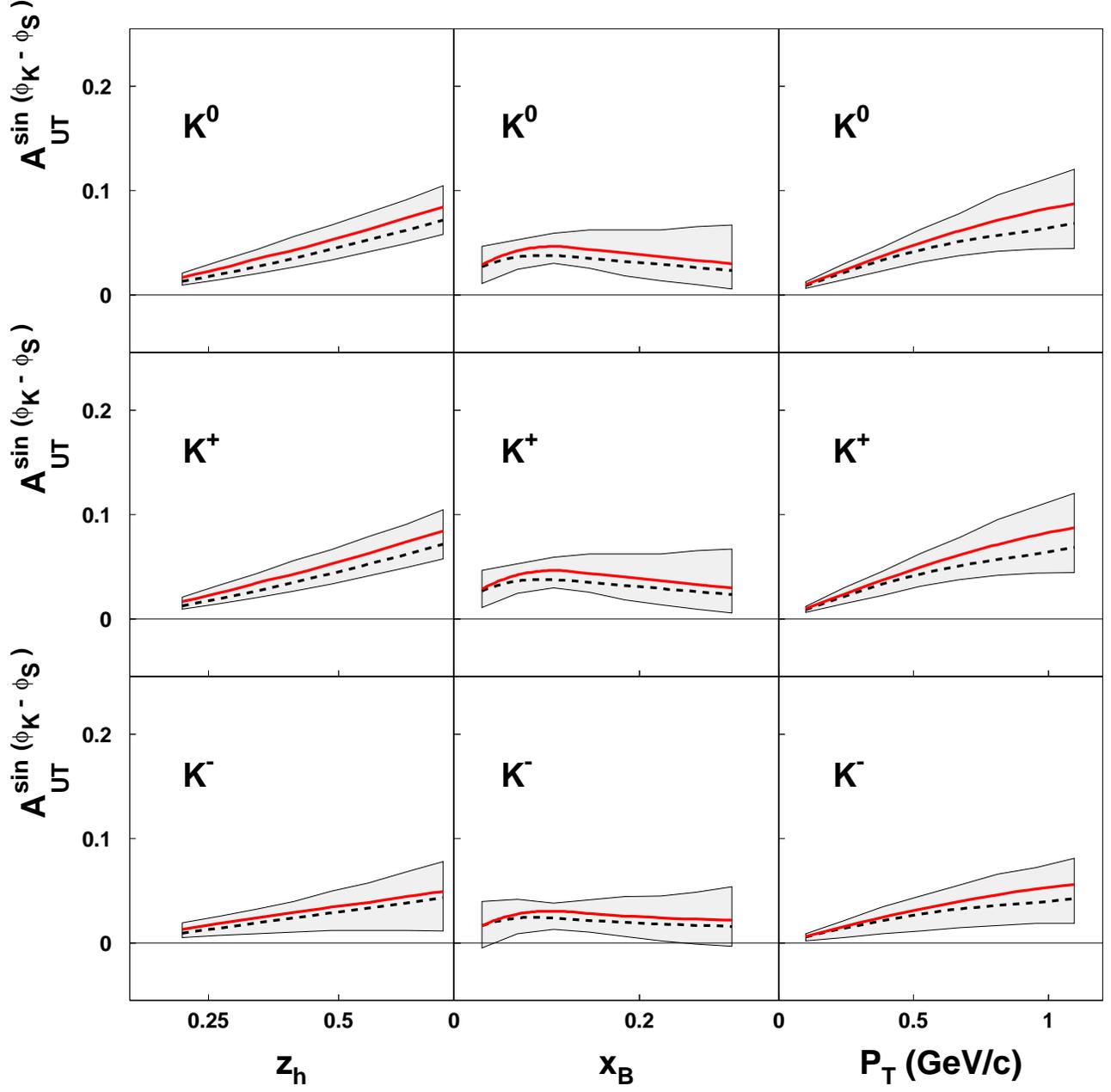} \caption{\small Predictions for  kaon
asymmetries $A_{UT}^{\sin(\phi_h-\phi_S)}$ at HERMES kinematics
for scattering off a transversely polarized proton target. The
curves correspond to calculation with exact kinematics (dashed
line) or keeping only terms up to ${\cal O}({\kt}/{Q})$ (solid
bold line). The shadowed region corresponds to the theoretical
uncertainty due to the parameter errors. \label{fig:authermesk}}
\end{center}
\end{figure}

\begin{figure}[T]
\begin{center}
\includegraphics[width=1.\textwidth,bb= 10 140 540 660]
{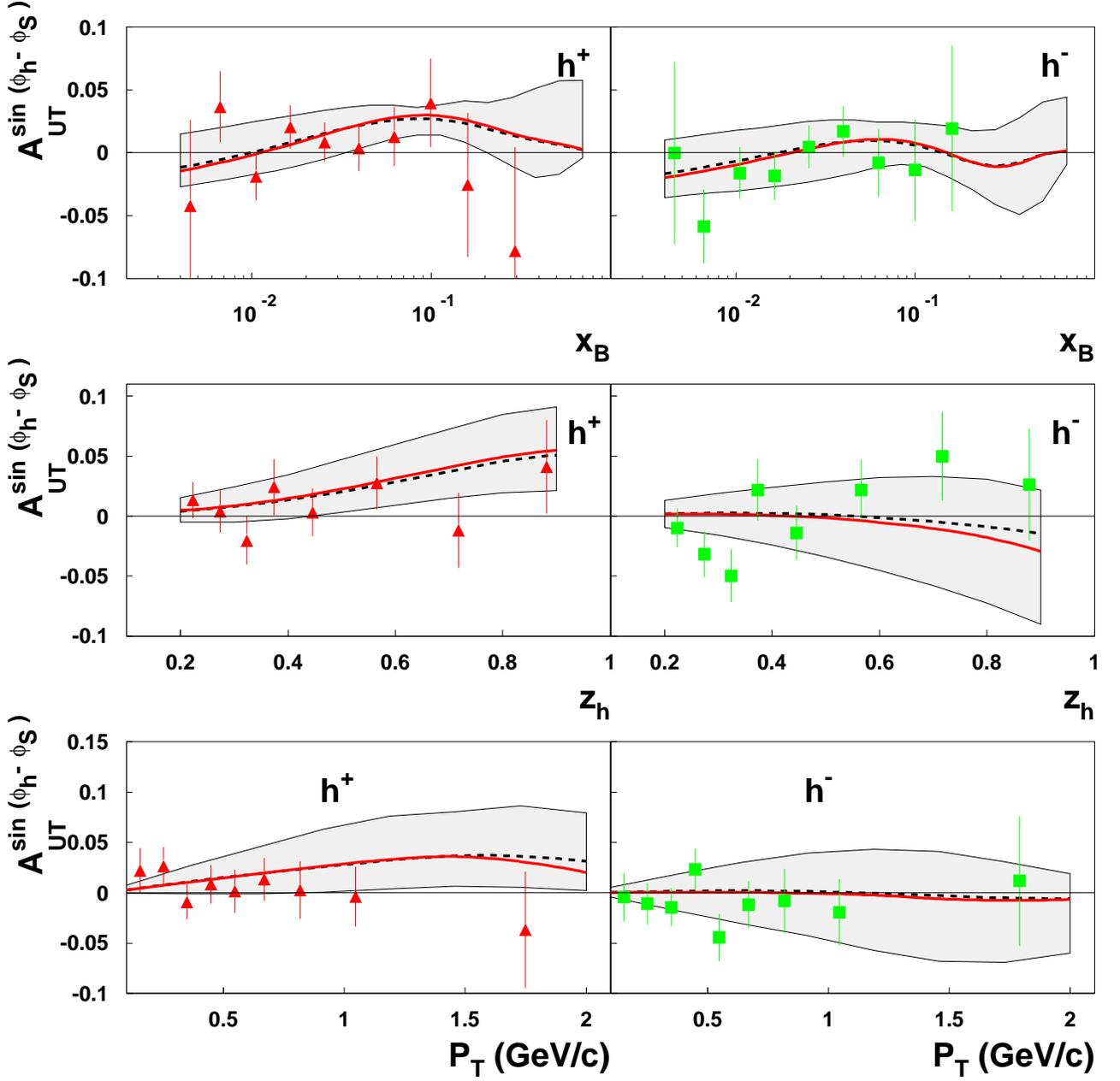}
\caption{\small
COMPASS preliminary data \cite{compUT} on $A_{UT}^{\sin(\phi_h-\phi_S)}$
for scattering
off a transversely polarized deuteron target and the production of positively
($h^+$) and negatively ($h^-$) charged hadrons. The curves show our
predictions, according to the values of the parameters for the Sivers
functions given in Table 1 and obtained from fitting the HERMES data on
$A_{UT}^{\sin(\phi_\pi-\phi_S)}$. Again, the dashed line refers to exact
kinematics, Eq. (\ref{hermesut}), while the solid bold line is obtained
by keeping only terms up to ${\cal O}({\kt}/{Q})$. The shadowed region shows
the theoretical uncertainty due to the parameter errors.}
\label{fig:autcompass}
\end{center}
\end{figure}

\begin{figure}[T]
\begin{center}
\includegraphics[width=1.\textwidth,bb= 10 140 540 660]
{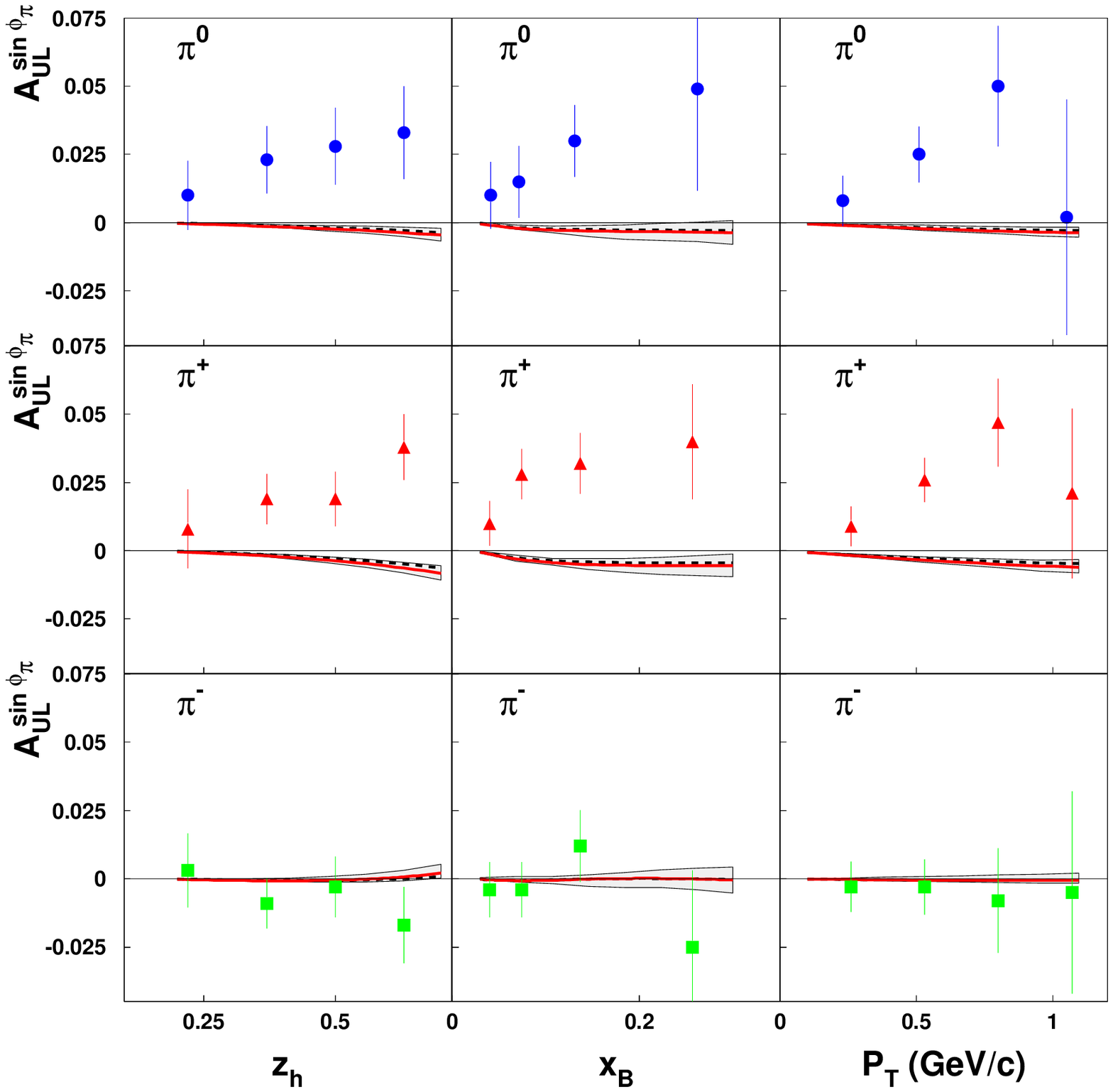}
\caption{\small
HERMES data on $A_{UL}^{\sin\phi_\pi}$ \cite{hermUL} for scattering off a
longitudinally polarized proton target and pion production. The curves show
the contribution of our Sivers functions alone, with exact kinematics
(Eq. (\ref{hermesul}), dashed line) or keeping only terms up to
${\cal O}({\kt}/{Q})$ (solid bold line). The shadowed region corresponds to
the theoretical uncertainty due to the parameter errors.}
\label{fig:aulhermes}
\end{center}
\end{figure}

We show the weighted SSA $A_{UT}^{\sin(\phi_\pi-\phi_S)}$  as a function of
one variable at a time, either $z_h$ or $\xbj$ or $P_T$; the integration over
the other variables has been performed consistently with the cuts of the
HERMES experiment, at $p_{lab} = 27.57$ GeV/$c$:
\bea
\nonumber
&& Q^2 > 1 \;  ({\rm GeV}/c)^2 \quad W^2 > 10 \; {\rm GeV}^2
\quad P_T > 0.05 \; {\rm GeV}/c \\
\label{hermutcuts}
&& 0.023 < \xbj < 0.4 \quad\quad 0.2 < z_h < 0.7 \quad\quad 0.1 < y < 0.85\>\\
&& 2 < E_h < 15 \> {\rm GeV} \nonumber.
\eea

A few comments are necessary for the interpretation of the results.
\begin{itemize}
\item
Fig.~\ref{fig:authermes} shows that a good agreement with experiments can
be obtained. However, due to the present quality of this first set of
data, the extracted Sivers functions are not well constrained and large
uncertainties are still possible.
\item
We notice that we have checked the compatibility of the HERMES data on
$A_{UT}^{\sin(\phi_\pi-\phi_S)}$ with the assumption of no Sivers effect,
$\Delta^N f \equiv 0$. The data show that the probability of a zero value for
 the Sivers function
 is less than $0.1$\%.
\item
It is interesting to compare the Sivers functions obtained here,
Eqs. (\ref{sivfunc}), (\ref{siversx}), (\ref{siverskt1}) and Table
1, with those obtained by fitting the SSA observed by the E704
Collaboration in $\pup \, p \to \pi \, X$ processes \cite{fu}. As
stressed in the Introduction the question regarding the
universality of the Sivers functions is a debated and open one.
The comparison of our results with Eqs. (46)-(48) of Ref.
\cite{fu} is not straightforward: one should keep in mind that
there is no sea contribution in Ref. \cite{fu} and that the HERMES
data are sensitive to much smaller $x$ values than the E704 ones
(which strongly depend on large $x$ values). Moreover, the average
$\kt$ and $p_\perp$ values adopted in Ref. \cite{fu} are somewhat
higher than those adopted here, derived with simplifying
assumptions from data on azimuthal dependences in SIDIS
processes. Despite all this, there is a clear indication that the
two sets of Sivers functions are not incompatible. The functions
of Ref. \cite{fu}, if used in our Eq. (\ref{hermesut}) or
(\ref{siversutapp})--(\ref{parsiversutapp}), still allow a
reasonable description of the HERMES data. On the other hand, our
Sivers functions of Table 1, if used to describe the SSA observed
by E704 experiment \cite{e704}, would overestimate the data at
small $x_F$ values; this could be easily corrected by gluon
contributions (gluon Sivers function) not considered in Ref.
\cite{fu} and absent at LO in SIDIS.

We do not wish, at this stage and with the limited amount of available
experimental information, to further stress such a point; the issue of the
uniqueness of the Sivers functions in SIDIS and $p \, p \to \pi \, X$
processes is still far from being phenomenologically established. We simply
conclude that it cannot be excluded by existing data.
\item
The Sivers functions obtained here are compatible with those extracted very
recently from an analysis of $P_T$ weighted HERMES data performed in
Ref. \cite{sivers_efremov}.
\end{itemize}

Leaving aside the question of the dependence of the Sivers functions on the
different physical processes, the consistency of our results can be checked
within SIDIS processes, by using our functions to give predictions for
other measured SSA. This can be done by computing, with our
sets of Sivers functions, the values of $A_{UT}^{\sin(\phi_\pi-\phi_S)}$
expected by the COMPASS experiment at CERN, which collects data in
$\mu d \to \mu h^\pm X$ processes at $p_{lab}=$ 160 GeV/$c$, spanning a
different kinematical region. Some preliminary results are already available
\cite{compUT}.  We neglect nuclear corrections and use the isospin symmetry
in order to obtain the parton distribution functions of the deuterium.
According to COMPASS experimental setup, we use the following cuts in the
numerator and denominator integration of Eq. (\ref{hermesut}):
\bea
\nonumber
Q^2 > 1 \;  ({\rm GeV}/c)^2 \quad W^2 > 25 \; {\rm GeV}^2 \quad
P_T > 0.1 \; {\rm GeV}/c \\
E_h > 4 \;{\rm GeV} \quad\quad
0.2 < z_h < 0.9 \quad\quad  0.1 < y < 0.9 \> .
\eea
The predictions of our model are presented in Fig.~\ref{fig:autcompass}
and compared with the available preliminary data \cite{compUT}. Within the
large errors, we find a good agreement, showing the consistency of the model.

We have also computed $A_{UT}^{\sin(\phi_K-\phi_S)}$ for kaon production,
$h = K$, which could be measured by HERMES; we have imposed the kinematical
cuts of Eq. (\ref{hermutcuts}), using the fragmentation functions given in
Ref. \cite{Kretzer}. Our results are given in Fig. \ref{fig:authermesk}.

Finally, we have considered the HERMES data on
$A_{UL}^{\sin\phi_\pi}$ obtained in the semi-inclusive
electro-production of pions on a longitudinally polarized hydrogen
target \cite{hermUL}. We have computed the Sivers contribution to
this quantity, according to Eq. (\ref{hermesul}), again with our
set of Sivers functions, and compared with data. Notice that no
agreement should be necessarily expected, as
$A_{UL}^{\sin\phi_\pi}$ can be originated also (even dominantly)
from the Collins mechanisms or higher-twist terms. Using the
following experimental cuts:
\bea
\nonumber
&& \hskip-18pt
1< Q^2 < 15 \;  ({\rm GeV}/c)^2 \quad W^2 > 4 \; {\rm GeV^2} \quad
P_T > 0.05 \; {\rm GeV}/c \quad 4.5 < E_h < 13.5  \; {\rm GeV} \\
&& \hskip-18pt
0.023 < \xbj < 0.4 \quad\quad 0.2 < z_h < 0.7 \quad\quad 0.2 < y < 0.85 \>,
\eea
we obtain the results depicted in Fig. \ref{fig:aulhermes}.

One can conclude that our Sivers functions extracted from the HERMES data on
$A_{UT}^{\sin(\phi_\pi-\phi_S)}$ only give a negligible contribution to
$A_{UL}^{\sin \phi_\pi}$. Not only, but the Sivers mechanism contributes
with opposite signs to the transverse and longitudinal SSA, as can be
seen from Eqs. (\ref{hermesul}) and (\ref{hermesut}). This implies that
Collins mechanism and/or higher-twist contributions are likely to be
wholly responsible for the observed  $A_{UL}^{\sin \phi_\pi}$, as suggested by
some authors \cite{c-s1}.

\section{Comments and conclusions}

We have studied inclusive and semi-inclusive DIS processes at leading 
order in the QCD parton model, in the $\gamma^* p$ c.m. frame and in the 
small $P_T \simeq \Lambda_{\rm QCD} \simeq \kt$ region, where intrinsic 
momenta dominate the final hadron azimuthal and $P_T$ distributions. We 
have adopted a factorized parton model scheme and exactly taken into account 
all intrinsic motions, of quarks inside the proton ($\bfk_\perp$) and of the 
final hadron with respect to the fragmenting quark ($\bfp_\perp$).

We have attempted a consistent treatment, assuming simple Gaussian $\kt$ and
$p_\perp$ distributions and extracting from various sets of SIDIS data
estimates about the average values $\langle \kt \rangle$ and
$\langle p_\perp \rangle$. Such values are assumed to be constant,
respectively in $x$ and $z$, and to be energy independent. Simple
parameterizations for the quark Sivers functions have been introduced.

The resulting picture has been applied to the computation of the weighted SSA
$A_{UT}^{\sin(\phi_\pi-\phi_S)}$, at LO in QCD parton model, which directly
depends on the intrinsic motions and the Sivers functions. The HERMES data 
clearly show a non zero Sivers effect; by a comparison
with these data some rough estimates of the Sivers functions for $u$ and
$d$ (both valence and sea) quarks have been obtained. These functions not
only describe well the HERMES data, but are also in agreement with some
COMPASS preliminary data on $A_{UT}^{\sin(\phi_\pi-\phi_S)}$, which refer
to different kinematical regions. The same functions are found to give
negligible contributions, with the wrong sign, to the measured longitudinal
SSA $A_{UL}^{\sin\phi_\pi}$. This asymmetry can indeed be originated by the
Collins mechanism and higher-twist contributions. Predictions for
$A_{UT}^{\sin(\phi_K-\phi_S)}$ for kaon production at HERMES have been given.

The quark Sivers functions extracted from the HERMES data on pion
$A_{UT}^{\sin(\phi_\pi-\phi_S)}$ have been compared with the Sivers functions
obtained by fitting the E704 data on SSA in $\pup \, p \to \pi \, X$
processes. Such a comparison cannot be considered as conclusive, as it refers
to situations with different kinematical regions and different assumptions
about the sea contribution; however, it does not exclude the possibility
that the two sets of Sivers functions -- those active in SIDIS and in $p\,p$
processes -- are the same. In particular the signs seem to be the same
in the two cases. Theoretical arguments support an opposite sign for
the Sivers functions in SIDIS and Drell-Yan processes, with no conclusions
concerning $p\,p$ interactions. Our Sivers functions are compatible with
those obtained in Ref. \cite{sivers_efremov}.

A phenomenological study of SSA and azimuthal dependences, within
a factorization scheme with unintegrated parton distribution and
fragmentation functions, is now possible. SIDIS processes with
measurements of the Cahn effect, and the various SSA
$A_{UL}^{\sin\phi_h}$, $A_{UT}^{\sin(\phi_h-\phi_S)}$ and
$A_{UT}^{\sin(\phi_h+\phi_S)}$ provide a rich ground to be further
explored, both theoretically and experimentally.

\begin{acknowledgments}
We would like to thank S. Gerassimov, D. Hasch, R. Joosten, P. Pagano and
G.~Schnell for enlightening discussions.
UD and FM acknowledge partial support by MIUR (Ministero dell'Istruzione,
dell'Universit\`a e della Ricerca) under Cofinanziamento PRIN 2003.
This research is part of the EU Integrated Infrastructure Initiative
HadronPhysics project, under contract number RII3-CT-2004-506078.
\end{acknowledgments}

\appendix
\section{}

It is known from symmetry principles that, within the one photon exchange
approximation, the double inclusive cross section for unpolarized SIDIS
processes, $\ell \, p \to \ell \, h \, X$, can have a dependence on the
azimuthal angle $\phi_h$ of the final hadron (in the reference frame
of Fig. 3) of the form \cite{aram}
\be
\frac{d^5\sigma^{\ell p \to \ell h X }}{d\xbj \, dQ^2 \, dz_h \, d^2 \bfP _T}
= A + B \cos\phi_h + C \cos2\phi_h \label{a1}
\ee
where $A, B$ and $C$ are scalar quantities which do not depend on
$\phi_h$. This is explicitly visible in the approximate expression
(\ref{cahn-anal-app}) and we wonder whether Eq.
(\ref{sidis-Xsec-final}) satisfies in general such a condition.

\mbox{}From Eqs. (\ref{stu}), (\ref{partond}), (\ref{partonf}), (\ref{zqh})
and (\ref{ptqh}) one can see that Eq. (\ref{sidis-Xsec-final}) is of the form
\be
\int d^2\bfk_\perp  \left[ a + b \, \bfl \cdot \bfk_\perp +
c\,(\bfl \cdot \bfk_\perp)^2 \right] \, F(\bfP_T \cdot \bfk_\perp, \dots)
\label{a2}
\ee
where $a, b$ and $c$ do not depend on angles and the $\dots$ stands for
scalar variables which also do not depend on azimuthal angles.

As a consequence, a tensorial analysis of Eq. (\ref{a2}) shows
that Eq. (\ref{sidis-Xsec-final}) can only contain azimuthal
dependences through the integrals:
\bea
&& b \> \bfl \cdot \int d^2\bfk_\perp \, \bfk_\perp  F(\bfP_T \cdot \bfk_\perp)
\sim  \bfl \cdot \bfP_T \sim \cos\phi_h \nonumber \\
&& c \> \ell_i \, \ell_j \int d^2\bfk_\perp \>
(\kt)_i \, (\kt)_j \, F(\bfP_T \cdot \bfk_\perp)
\sim \ell_i \, \ell_j \, (P_T)_i \, (P_T)_j \sim \cos^2\phi_h =
\frac{1 + \cos2\phi_h}{2}
\nonumber
\eea
which agree with Eq. (\ref{a1}).

\end{document}